# Metal Oxide-based Gas Sensor Array for the VOCs Analysis in Complex Mixtures using Machine Learning

**Shivam Singh[1], Sajana S[1], Poornima Varma[2], Gajje Sreelekha[3], Chandranath Adak[3,*], Rajendra P. Shukla[4,*], Vinayak B. Kamble[1,*]**

[1]School of Physics, Indian Institute of Science Education and Research Thiruvananthapuram, 695551 India.

[2]Dept. of CSE, Indian Institute of Information Technology Lucknow, Uttar Pradesh 226002, India.

[3]Dept. of CSE, Indian Institute of Technology Patna, Bihar 801106, India.

[4]BIOS Lab-on-a-Chip Group, MESA+ Institute for Nanotechnology, Max Planck Center for Complex Fluid Dynamics, University of Twente, P.O. Box 217, 7500 AE Enschede, The Netherlands.

Corresponding authors: Chandranath Adak, Rajendra P. Shukla, Vinayak Kamble.

**Abstract**

Detection of Volatile Organic Compounds (VOCs) from the breath is becoming a viable route for the early detection of diseases non-invasively. This paper presents a sensor array of 3 component metal oxides that give maximal cross-sensitivity and can successfully use machine learning methods to identify four distinct VOCs in a mixture. The metal oxide sensor array comprises NiO-Au (ohmic), CuO-Au (Schottky), and ZnO –Au (Schottky) sensors made by the DC sputtering method and having a thickness of 80-100 nm. The NiO CuO films have ultrafine particle sizes of

<50 nm and rough surface texture, while ZnO films consist of nanosheets. This array was subjected to various VOC concentrations, including ethanol, acetone, toluene, and chloroform gases, one by one and in a pair/mix of gases. Thus, the response values show drastic interference and departure from commonly observed power law behavior. The dataset obtained from individual gases and their mixtures were analyzed using multiple machine learning algorithms, such as Random Forest (RF), K-Nearest Neighbor (KNN), Decision Tree, Linear Regression, Logistic Regression, Naive Bayes, Linear Discriminant Analysis, Artificial Neural Network, and Support Vector Machine. KNN and RF have shown more than 99% accuracy in classifying different varying chemicals in the gas mixtures. In regression analysis, KNN has delivered the best results with $R^2$ value of more than 0.99 and LOD of 0.012, 0.015, 0.014 and 0.025 PPM for predicting the concentrations of varying chemicals Acetone, Toluene, Ethanol, and Chloroform, respectively in complex mixtures. Therefore, it is demonstrated that the array utilizing the provided algorithms can classify and predict the concentrations of the four gases simultaneously for disease diagnosis and treatment monitoring.

**Keywords:** Gas sensor, sensor array, Metal oxides, VOCs, Complex mixture, Machine learning.

# 1. Introduction

With the advent of modern technology, there is much interest in reducing surgical involvement and enhanced early identification of illness. Since it is quicker, less intrusive, and more accessible than a traditional clinical assessment, identifying certain illnesses employing human exhaled air has garnered great interest[1]. In this context, exhaled breath is the ideal non-invasive approach since it accurately captures the metabolic processes occurring within the human body[2]. Compared to lab tests, disease identification utilizing expiratory VOCs has emerged as the preferable approach for early screening. Besides, it has another excellent relevance for continuous breath monitoring for knowing health anomalies that appear transient or periodic. Breath monitoring has several benefits[3], the most significant among them being a simple, quick, and straightforward sampling collection method provided by its non-invasive approach[4].

Many (about hundreds) of volatile chemical molecules are found in an individual's breath. Some Volatile Organic Compounds (VOC) chemicals, notably isoprene (heart disease), acetone (diabetes), toluene (lung cancer), nitrogen monoxide (asthma), pentane (heart disease) and ammonia (kidney dysfunction) are established indicators that anticipate underlying disorders. However, several variables affect the constitution of exhaled breath and can be broadly classified as lifestyle-based, health-based, and environment-based. The usual range for toxicants in a person's exhaled breath is between parts per billion (PPB) to parts per trillion (PPT)[5]. The number of VOCs and their relative proportions are specific to the health of individuals, or unexpected VOCs may be released by irregular metabolic reactions[6]. Therefore, breath evaluation is often used to identify various diseases, including renal dysfunction, prostate cancer, and other types of cancers[7]. Identifying various indicators for each ailment makes it possible to distinguish between healthy people and those with illnesses using a sensor array. It is also possible to continually monitor those using wearable technology[8].

The brief involvement of these VOCs in various diseases through exhalation and their severe effect on the human body are expressed in detail in the online resource file section 1. We have identified those common VOCs like ethanol, toluene, acetone, and chloroform, among the biomarkers routinely used to analyze the response.

Current-state-of-the-art technologies use gas chromatography followed by mass spectroscopy to analyze breath samples to investigate specific VOCs in patient samples. Although those are precise, these techniques require a sophisticated setup and trained individuals to handle those, increasing the analysis cost. Moreover, it is also a time-consuming process to get the analysis report from centralized laboratories. These techniques also use labeling or pretreatment of the samples, which may affect the exact levels of the VOCs in complex media.

Recently, chemiresistive gas sensors have been explored to analyze VOCs in breath samples due to their simple design, high sensitivity, fast response time, and cost-effectiveness[9]. These sensors can be employed at the point-of-care for VOC analysis. Metal-oxide-based gas sensors have gained significant interest in these sensors due to their small size, ease of operation, inexpensiveness, excellent sensing performance, and low maintenance. However, despite the high sensitivity and fast response time, these sensors have yet to reach clinical studies due to the presence of interfering species generating overlapping and masking gas-sensing signals. The electrical signals generated from the gas sensor in a multi-component mixture solution are difficult to differentiate between signals of the target analyte and interfering species. In the recent past, another approach called "electronic nose" where a gas sensor array has been utilized in place of a single sensor to record the response in a multi-component mixture solution, and the data was analyzed using Machine Learning (ML) algorithms[10].

The gas sensor array consists of non-specific sensors in the array and records the fingerprints of the multi-component mixture solution. This approach reduced the effect of interfering species and required no pretreatment of the breath samples, thereby shifting the challenges of gas sensing from the physical to the digital domain.

Thus, the work presented in the manuscript projects the metal oxide sensor array for diagnostics of various biomarkers from breath for early detection of diseases. However, the biomarkers in breath are exhaled as mixtures, and therefore their identification as well as quantification is a challenging task. Here, we use machine learning methods for identification (classification) and quantification (regression). Besides, the sensor array is made of reliable metal oxide thin films fabricated using a sputtering method. Therefore, not only qualitative but quantitative detection of four VOCs simultaneously allows the detection of multiple diseases and monitoring of the health of individuals.

# 2. Experimental Details

In this section, we discuss the fabrication the of MOS gas sensor array followed by experiment setups for ML-based gaseous chemical classification and regression analysis.

## 2.1. Fabricating metal oxide (MOS) gas sensor array

### 2.1.1. Thin film deposition using DC-RF magnetron sputtering

DC reactive magnetron sputtering was used to create thin films of CuO, NiO and ZnO onto both glass and alumina substrates. These interdigitated gold electrodes (IDE) equipped alumina substrates comprise two finely constructed, closely spaced gold electrodes and two connecting tracks that have all been diligently sculpted. It had a ceramic substrate that was 22.8 x 7.6 x 1 mm in length, width, and height. Notably, there was a distinct gap of 200µm between the interdigitated gold electrodes. During DC magnetron sputtering, the metal (Copper, Nickel, and Zinc) targets (99.99%) of 1 inch in diameter and a few millimeters thick were employed. The sputter gas was pure argon (99.9997%), while the reactive gas was pure oxygen (99.9997%). Mass flow regulators controlled both gas flows independently. The sputtering chamber was vacuumed to a base pressure of about $10^{-6}$ mbar with the help of a turbo molecular vacuum pump and a rotary mechanical backing pump before the thin oxide films were deposited. The input parameters of different voltages and currents were used. The constant Argon flow rate was 30 SCCM. Pre-sputtering was kept going for 10 minutes to ensure the target surface was thoroughly scrubbed. Following the pre-sputtering step, 10 SCCM of oxygen was added into the reaction chamber while the deposition pressure was maintained at a constant ~ $10^{-2}$ mbar. Thin film deposition on substrates may begin once the shutter is opened. The optimum deposition time ($t_d$) was different for three oxides, while the optimum substrate temperature (Ts) was 300 K. After rotating the substrates while maintaining a distance of 6-8 cm from the target, we found the best results. Table S1 shows the variation of deposition parameters for all three oxide films. The sputtered samples are shown digitally in ESM_1(inset).

#### 2.1.2. Material characterization of MOS gas sensor array

Powder X-ray diffraction (XRD) was used to examine the microstructure and crystallinity of materials using a Bruker Powder XRD device utilizing Cu kα radiation (λ = 1.5418 Å) and a nickel filter. Data were gathered at a scan rate of 2 data points per minute, with steps of 2 thetas ranging from 10 to 80 degrees. The films' surface morphology was captured using a Nova NANOSEM 450 equipped with WDS and EDS. We used a secondary emission mode with an operating voltage of 15 kV for this particular picture capture. EDS was used to verify the composition of the elements.

Raman spectroscopy was performed with a Horiba scientific Xplora plus spectrometer using a 514-nanometer-wavelength argon laser. The samples' thickness was determined using a KLA Tencor D600 stylus surface profiler equipped with a step height measuring system.

### 2.1.3. The electrical measurements

The I-V characteristics of the oxide thin films deposited in the alumina substrate with gold IDEs were investigated from room temperature to 300 $^{o}$C. The R vs. T measurements reported here are done in an equilibrium fashion. i.e., the temperature is held constant at the desired values (within the experimental limits ±1$^{o}$C, the heating stage made by Linkam, UK), and the sample is thermalized at a given temperature. The system had a platinum (PT100) temperature measurement and control sensor. The system was under ambient conditions, as the final sensing studies would be done in ambient conditions. The bias voltage was swept between -10 V and +10 V to each sensor at ambient temperatures. The resistance values at each temperature were calculated from I-V slopes.

### 2.1.4 Gas sensing studies using MOS sensor array: Experimental setup

Gas sensing experiments were carried out by observing how the thin films' electrical resistance changed in response to various VOCs at fixed operating temperatures. The sample gases were infused under dynamic flow conditions fixed by mass flow controllers (Maker Alicat, United States) with varying capacities. The details of the vapor concentration calculation are in the Online Resource section.

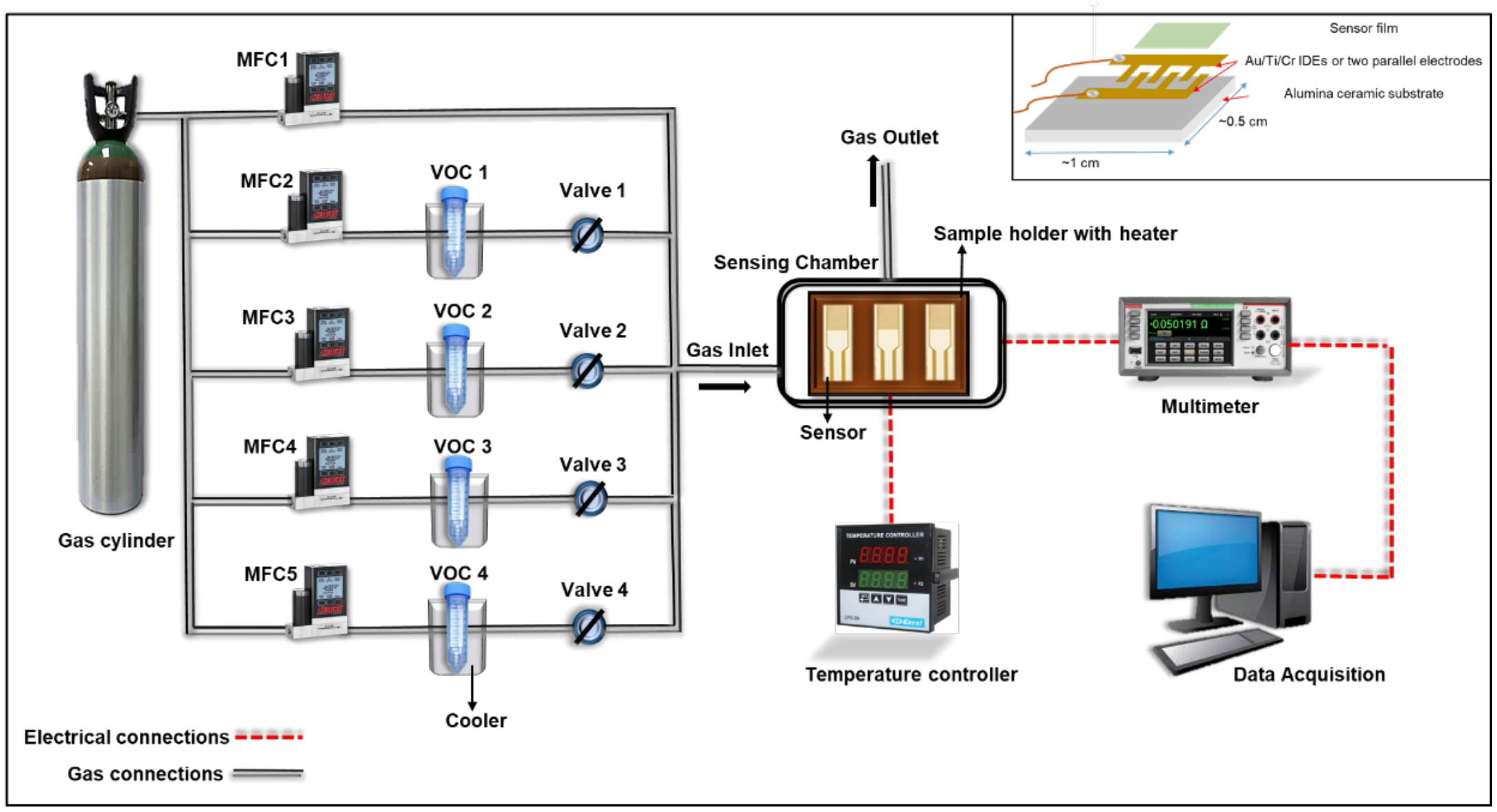

Fig 1: The Schematic diagram of the gas sensing setup used for the experiment.

Fig 1 depicts the indigenous gas sensing system used in this investigation. The films were mounted on a sample holder that could reach 400°C using a heater underneath the sample holder to evaluate the gas sensing characteristic in a detecting chamber (make Excel Instruments). A type-K thermocouple-inserted sample holder was used to measure the sensor's temperature. An Alumina substrate with interdigitated Gold electrodes is utilized for measuring sensor resistance. The details of the substrate and electrode dimensions are in the Online Resource section. A Keithley 6517B electrometer linked to a workstation was used to measure the sensor resistance by applying a consistent bias voltage of 10 V to two probes. With a tolerance of 1 fA, it's a high-resistance analyzer that could contribute meaningfully to $10^{15}$ ohms. Exposing the deposited films to the appropriate vapors diluted in air was necessary to test the sensor response to ethanol and other volatile organic chemicals. The % response (S) was calculated using eq (1) as below.

$$\% \, Response, S = \frac{Ra - Rg}{Ra} \qquad (1)$$

Where Ra is the sensor resistance within airflow and Rg is the sensor resistance when the test gas is present. It should be noted that the response sign for n-type and p-type devices is opposite to the stated gas. The sensor resistance decreases when n-type material is exposed to a reducing gas because the gas injects excess carriers into the material. But as the resistance decreases, the resistance changes the most, increasing 100% monotonically. However, suppose the resistance increases due to gas exposure, as in the case of p-type material subjected to reducing gas. In that situation, the standard deviation of the resistance change is more significant than 100% or greater than double the original value.

The chemiresistive array sensing tests were done by passing a set amount of target gas mixed with a predefined proportion of air, determined by the equalization method at fixed intervals. Both with and without the analyte, the total flow was maintained at 500 SCCM, and the two-probe mode was used to collect the sensor's resistance data. The sensors were tested by being exposed to ethanol concentrations of 100–2400 ppm at 200 °C. Individual response research utilizing toluene, chloroform, and acetone was also carried out under identical conditions. By cooling the liquids in the tube to the same temperature and using the same MFC dispersion ratios, similar studies were conducted at 200 °C to examine ethanol's cross-sensitivity to other gases such as toluene, chloroform, and acetone. The volatile liquids are used to generate the vapors of the desired gas for sensing. Here, the bubblers are maintained at a constant temperature, and the carrier gas is bubbled through the liquid in the thermostat to generate the vapors subjected to sensor exposure. Here, the concentrations of the vapors are mainly governed by constant temperature baths and the flow rate of the carrier gas to a certain extent. Therefore, the gas concentrations utilized were primarily governed by the generation rate and vapor pressure.

## 2.2. Gaseous chemical classification and regression analysis using machine learning models

The dataset used in this study consists of gas sensor data comprising three different mixtures. Each mixture represents a distinct scenario based on the number of gases present, namely

i. *1-gas:* a single gaseous chemical,

ii. *2-gases:* mixture with one constant chemical and one varying chemical,

iii. *3-gases:* mixture with two constant chemicals and one varying chemical.

The chemicals involved in these mixtures are Acetone, Toluene, Chloroform, and Ethanol. We performed the analysis on these 1-gas, 2-gases, and 3-gases datasets. The primary objective was to explore and investigate the region where the concentrations of these interfering biomarkers were high. Subsequently, the sensor response was recorded by introducing variable gas concentrations within this specific range, as mentioned above. This approach thoroughly examines and characterizes the sensor's behavior when exposed to various interfering gases at different concentrations. Besides, the gas sensing apparatus' practical limits, such as MFC resolution accuracy, primarily determined our study's interference gas concentration selection. We have been focusing on our system's capacity to manage intricate gas combinations while optimizing and miniaturizing them. Working with metal oxides has been a critical component of our strategy since it allows us to extrapolate response concentration curves to lower parts per million (ppm) concentrations. Tapping into the power of machine learning to improve our system's accuracy and forecasting powers at these lower concentrations is envisaged. The strategy is to be rigorously trained and fine-tuned to produce highly accurate predictions even at sub-ppm levels. All possible combinations of four biomarkers were employed to record the readings for the mixtures with two and three gases, mentioned in Table S2. A gaseous chemical that needs to be classified or whose concentration needs to be anticipated is kept variable for datasets with mixtures of gases. Three sensing components (CuO, NiO, and ZnO) are used in the dataset to record measurements. The objective is to predict or classify the concentration of the varying gas, whether it is Acetone, Toluene, Chloroform, or Ethanol.

Each dataset has the following sample rows: 2241436 sample rows for 1-gas, 227617 rows for the mixture of 2-gases, and 131120 rows for the mixture of 3-gases. There are 6, 8, and 10 columns in the abovementioned datasets. The correlation matrices for the three datasets are shown in ESM_3 in Online Resource. We performed two types of analysis: (i) *classification* to categorize the varying gas/ chemical and (ii) *regression* analysis to predict the concentration of the gas.

For the classification task with the 1-gas dataset, we used five features, i.e., resistance, time, concentration in terms of parts per million (PPM), temperature, and electrode, to categorize the varying chemicals. For the classification task with 2-gases dataset, we used seven features, i.e., time, ZnO_resistance, NiO_resistance, CuO resistance, constant_chemical (CC), CC_PPM and varying_chemical_PPM (VC_PPM) to classify varying_chemical (VC). For 3-gases dataset, we used nine features, i.e., time, ZnO_resistance, NiO_resistance, CuO resistance, constant_chemical_1 (CC_1), CC_1_PPM, CC_2, CC_2_PPM, and varying_chemical PPM to classify varying_chemical (VC). For regression tasks with 1-gas, we predicted the gas concentration in PPM; for both of the 2 and 3 gases datasets, we predicted VC_PPM. The rest of the column values were used as features. For the experimental analysis, the dataset was divided into training, validation, and testing sets with a ratio of 56:14:30. To assess the performance of the classification analysis, the accuracy metric was used; and for the regression analysis, the mean absolute error (MAE), mean squared error (MSE), root mean square error (RMSE), normalized

RMSE (NRMSE), coefficient of determination ($R^2$), Limits of Detection (LoD), and Limit of Quantification (LoQ) was employed[11, 12]. Here, we present the results of the test dataset.

We observed significant outliers in the dataset. Outliers in the input data may distort and deceive ML models during training, leading to longer training times, less accurate models, and ultimately worse outcomes. Therefore, the outliers were eliminated using the data quantile information[13] defining an upper and lower limit. A data value was eliminated from our primary data frame if it exceeded the upper limit or fell below the lower limit. The datasets underwent preprocessing steps to conduct a comprehensive analysis, including outlier detection and removal, min-max scaling to handle variations in feature values, and label encoding to address categorical features[14]. Categorical data[15] was encoded using label encoding, as only eight distinct values were in the categorical column. It is crucial to convert categorical data into a numerical format to enable processing by ML models. Other approaches for categorical data include one-hot encoding, vectorization, and label encoding. Upon completing the dataset preprocessing, models were built using the selected algorithms. Rigorous hyperparameter tuning was performed for all the algorithms employed in this gas sensor dataset analysis. Grid search cross-validation was utilized for hyperparameter tuning[16].

***Machine configuration***: All the ML-based analyses were performed on the TensorFlow-2 framework having Python 3.7.13 over a computer with Intel(R) Xeon(R) CPU @ 2.00GHz having 52 GB RAM and Tesla T4 16 GB GPU.

# 3. Results

This section discusses the fabrication of devices, characterization of the gas sensor array, ML-based classification, and regression of gaseous chemicals.

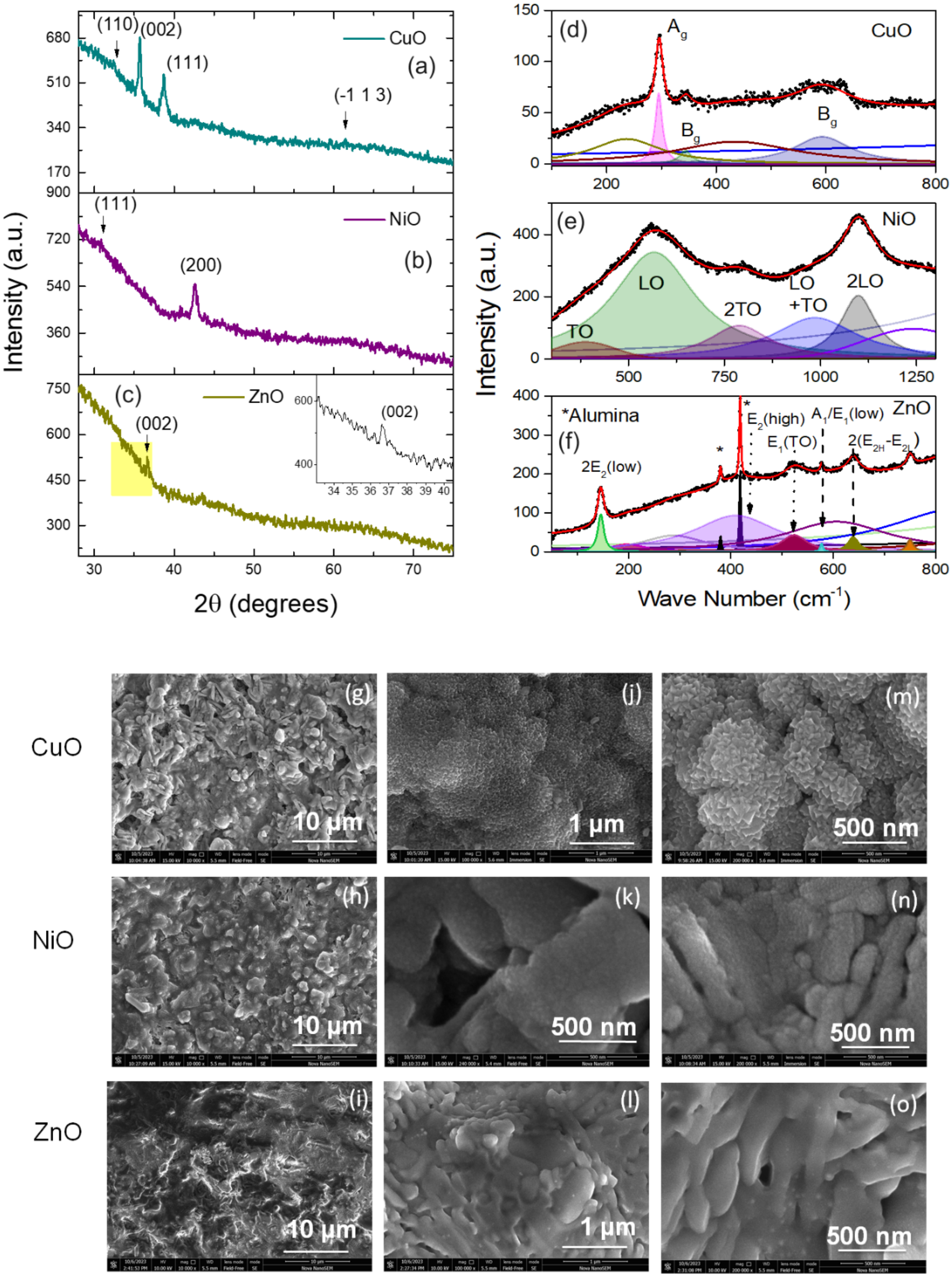

Fig. 2: The XRD patterns and Raman spectra of (a, d) CuO, (b, e) NiO, and (c, f) ZnO thin films at room temperature (for XRD, the samples were also deposited on glass substrates). The Scanning electron micrographs of CuO, NiO, and ZnO at low (g-i) and high (j-l) magnifications, respectively. EDS spectra of (m) CuO, (n) NiO and (o) ZnO.

## 3.1. Device fabrication, characterization of the gas sensor array and sensing studies

To create our device, we used a DC reactive magnetron sputtering technique to deposit copper, nickel, and zinc oxide on an alumina substrate having interdigitated gold electrodes with the corresponding metal targets. Table S1 represents the sputtering parameter to ensure the deposition process is accurate. Pre-deposition of 10 minutes was done to ensure that the surface was thoroughly scrubbed and no contamination was left. ESM_1 (inset) represents the schematic diagram of our fabricated device. The gold electrodes were vital because they helped the device detect resistance changes when exposed to different gases.

### 3.1.1. Material characterization of MOS gas sensor array

The as-prepared oxide thin films were deployed for the sensor array, and the same were examined using various techniques to explore the particular details of the structure and composition. Although the samples used in gas sensing are deposited on Alumina substrates, the XRD of those films was primarily dominated by highly crystalline Alumina substrate peaks. Therefore, to confirm the crystallinity of each synthesized sensor film, the same was also deposited on glass and investigated using XRD. The corresponding XRD patterns of CuO, NiO, and ZnO are displayed in Fig. 2(a, b, and c). From all these XRD data, it may be inferred that each of the oxide layers is formed albeit with a thickness estimated at 80-100 nm, which results in the poor intensity of peaks. No peaks corresponding to any impurity are seen to the best of the resolution in any of the XRD patterns. ZnO and NiO show a firm texture in crystallinity marked by a single diffraction peak. This implies that the films are preferentially oriented (except CuO) along a certain direction[22]. This happens due to homogeneous nucleation of the oxide crystals, which grow along the crystal's energetically most favorable (lowest formation energy) planes. Moreover, the broadening of the peaks reflects a smaller crystallite size, possibly due to a lack of energy for long-range growth as the deposition is carried out at room temperature. Nevertheless, such small crystallite size and low thickness are favorable for gas sensing as the sensing response is dramatically improved if the dimensions are of the order of space charge region[17]. Along with the crystalline structure, the morphology (shape, grain size, porosity, etc.) of the sensor films significantly affects the sensing attributes of the chemiresistive sensors. Therefore, the microstructure and morphology of the films are examined using scanning electron microscopy and microscopic composition analysis using energy-dispersive X-ray spectroscopy. Fig. 2(g-l) shows the same for all three films at low and high magnifications.

**Copper oxide:**

The Fig. 2(a) illustrates the CuO XRD data with characteristic patterns for the (002), (-111), and (111) reflections of a monoclinic CuO with lattice constants of 5.13 Å, 3.42 Å, and 4.68 Å (JCPDS: 01-073-6023). Because of the films' low thickness (~80 nm), the XRD pattern is not significant in analyzing the crystallinity of the films. Therefore, Raman spectra have been investigated for all three samples at room temperature. Here, the signal is collected from the tiny focus of the laser beam on the film surface and is organized in back reflection geometry. Therefore, it shows much better sensitivity to the surface than the substrate that lies underneath. The Raman spectra identify their vibrational properties at ambient temperature and are found to give peaks that are unique to each material. The copper oxide Raman spectrum on an alumina substrate is shown in Fig. 2(d), depicting Raman modes at 292.6, 345.3, and 627.5 $cm^{-1}$. The positions of the peaks in the spectra with this specimen are in close vicinity of those corresponding reported CuO values[18, 19]. Besides, the peak at low energy, i.e., 224.9 $cm^{-1,}$ could be due to local partial suboxide, as seen by Debbichi et. al[20]. Several factors, such as poor crystallinity, an accumulation of structural faults in the crystalline lattice, and fluorescence of the incident radiation, may be responsible for the broad baseline around 100 and 800 $cm^{-1}$ seen in this spectral region.

When observed under a scanning electron microscope, the films look rough in texture and are coated onto alumina grains uniformly (Fig. 2(g)). The finer crystallite size is seen distinctly in a scanning electron microscope at high magnification (Fig. 2(j and m)). It should be noted that the films are deposited on a polycrystalline alumina substrate that has a distinct grain structure of particle size nearly 2-3 μm. The same is seen in SEM images of all the films. However, the sensing oxide film deposited on its top takes an almost conformal shape of the alumina substrate grains. Fig. 2(m) shows the finer crystallite size, having distinct triangular morphology. In the EDS spectrum shown in ESM_4(a) in the Online Resource section, the film shows only copper, oxygen, and Aluminum from the Alumina substrate. The quantification is challenging because of the oxygen signal from the bottom oxide substrate.

**Nickel Oxide:**

The (111) and (200), considered as the top of NiO (JCPDS: 00-047-1049), were matched by the middle XRD pattern, which corresponds to a cubic arrangement with lattice constants of **a** = **b** = **c** = 4.17 Å. Fig. 2(e) depicts the Raman spectra recorded for NiO thin films deposited on a glass substrate for 18 minutes. As per identification in ref[21], the observed peaks may be ascribed to the one-phonon constituting TO (389.2 $cm^{-1}$) and LO (567.7 $cm^{-1}$) modes. The second harmonic, i.e., 2TO (789.5 $cm^{-1}$) and 2LO modes (1099.5 $cm^{-1}$), confirms the phase. The Ni-O bond's stretching mode and flaws are both indicated by the peak LO's considerable breadth (576.7 $cm^{-1}$)[22, 23]. The broad nature of the peak arises from the finer crystallite sizes, and therefore, a significant overlap exists among the peaks.

As discussed in the case of copper oxide, it may be seen that the particles of nickel oxide film are also clustered, making it rougher in texture, as seen in Fig. 2(h). In this case, several open pores have diameters of 500 nm. The film develops in tiers, and the texture appears granular (see Fig.s 2(k and n)). Nevertheless, such a high surface roughness and, thereby, high surface area benefits the gas sensing devices. In this case, the chemical composition examined using the EDS spectrum shows only Ni and O other than the Al signal contributed by the substrate, as shown in ESM_4(b).

**Zinc Oxide:**

The bottom-most XRD pattern, i.e., Fig. 2(c), shows a single prominent peak that matches the ZnO wurtzite structure for (002) reflection (JCPDS: 00-036-1451), and it possesses a hexagonal structure with cell parameters (**a** = **b** = 3.25 Å and **c** = 5.21 Å). The Raman modes $A_{1g}$(TO) positioned at 380.38 $cm^{-1}$ $E_2$(H) at 418.7 $cm^{-1}$ constitute the vibrational configurations corresponding to the hexagonal wurtzite geometry of ZnO[24, 25] on an alumina surface, as shown in Fig. 2(f). However, the peak at 321.0 $cm^{-1}$ matches up to the second-order vibration mode originating from the zone boundary phonons [E2(high)–E2 (low)] of hexagonal ZnO[26]. Besides, two broad peaks contribute to the extensive background, which could be due to the finite size of the crystallites and the broadening mentioned earlier. Overall, the three samples' Raman spectra show very low intensities and significant peak broadening. Like XRD, this broadening results from the sufficiently small size of the crystallites. Therefore, these results are in good agreement with that of the XRD of the films. However, it shows better confirmation of single-phase oxide films and their nanocrystalline nature.

Unlike the other two films, the ZnO film imaging was challenging due to the excessive charging effect due to the poor conductivity. Typically, at high magnification (Fig.2(l and o)), ZnO films are lamellar structures that are further composed of smaller grains. The film showed a 2D sheet-like nature towards the edges, and the presence of sheets and (002) seen in Fig 2 (i and l), a single peak in XRD, points to the same. The same could also be the reason for the sharp peak in the Raman spectrum. Besides, the trigonal structure of alumina could help grow 2D sheets of ZnO that have hexagonal symmetry in the c-plane. The EDS spectra consist of Zn and O elements (except Al arising from the substrate). More SEM micrographs are shown in Fig S4 in the Online Resource.

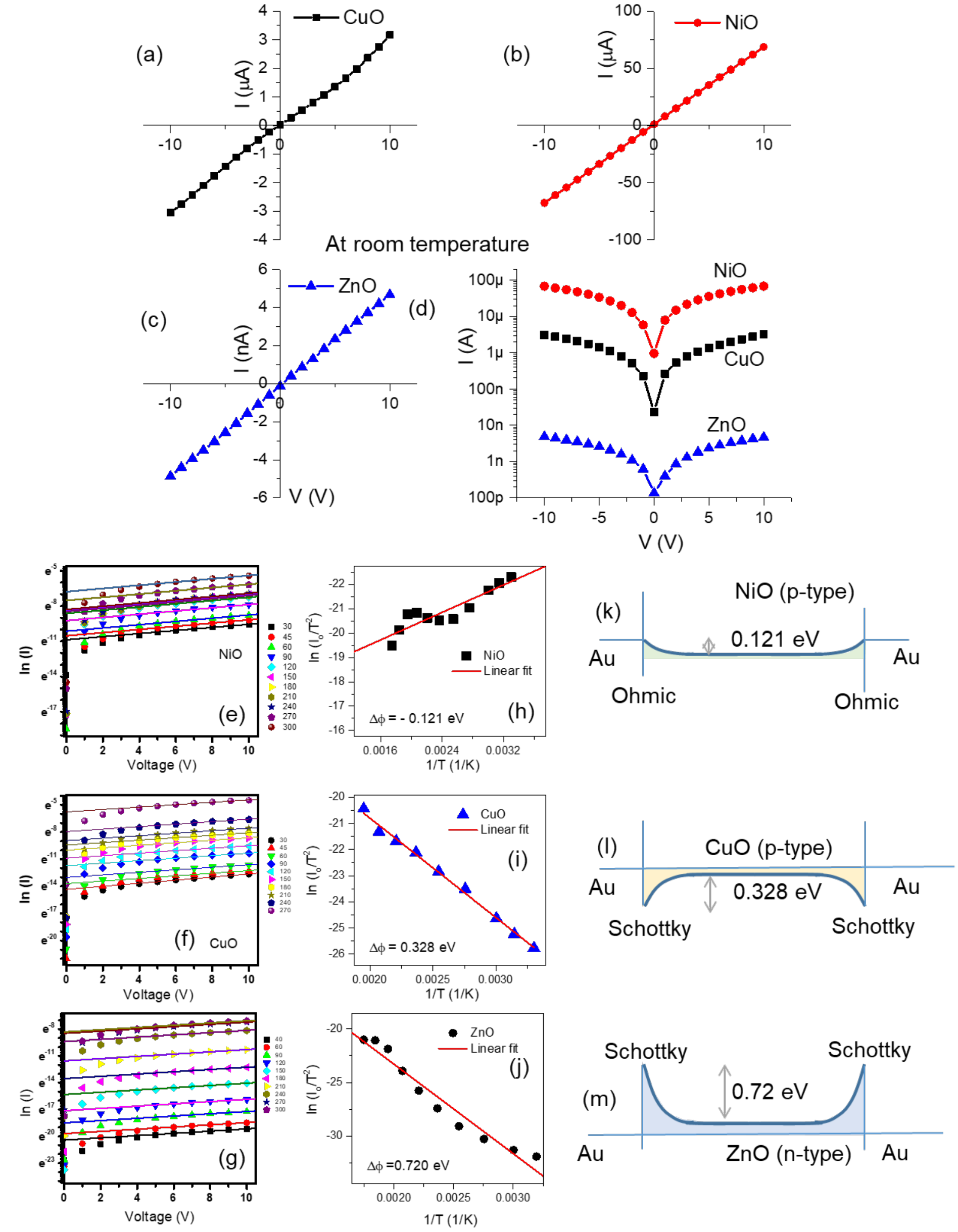

Fig 3. The room temperature I-V of (a) CuO, (b) NiO, (c) ZnO and (d) their comparison together. The thermionic emission model analysis of temperature dependent I-V data and $\ln(I_o/T^2)$ vs 1/T plot for estimating the barrier height for (e, h) NiO. (f, i) CuO and (g, j) ZnO thin films with Au electrodes. The corresponding band bending is shown in (k) NiO, (l)CuO and (m) ZnO.

### 3.1.2. Electrical Measurement of the MOS gas sensor array

The I-V characteristics of the sensors were measured from room temperature up to 300°C. The I-Vs at room temperature are shown in Fig. 3 (a-d), while the entire temperature range is shown in ESM_5(a, d), S5(b, e), and S5(c, f) for CuO, NiO, and ZnO. The resistance values so deduced were plotted as a function of temperature, and all the samples demonstrated a typical insulating/semiconducting nature. (See ESM_6 in the Online Resource section). The typical value of resistances was about 500 kΩ, 20 MΩ, and 100 MΩ for CuO, Ni, O, and ZnO, respectively, at room temperature. These dropped to 423 Ω, 14 kΩ, and 684 kΩ at 300°C for CuO, NiO, and ZnO, respectively.

The IVs were mainly linear. However, some showed a slight nonlinearity, such as CuO, mostly at room temperature (see Fig. 3(a)). To explore the nature of the contact, the I-Vs were analyzed using a thermionic emission model wherein the temperature dependence of current followed the Richardson –Dushman equation (2)[27, 28],

$$I = SA^{*}T^{2}e^{-\frac{e\phi}{k_B T}}\left(1 - e^{\frac{qV}{k_B T}}\right) \qquad (2)$$

Here, $k_B$ is Boltzmann constant, $A*$ is Richardson constant and $S$ represents the device area. It may be shown that following the same equation, the slope of the graph of $ln(I/T^2)$ vs 1/T allows estimation of the barrier height, which is the difference between the work function (WF) of the metal and semiconductor. Knowing the WF of the metal (5.1 eV for Au), it is easy to deduce the same of the oxide semiconductor. The same plots for all three oxides are shown in Fig. 3(e-g and h-j). Of the three, only NiO has formed ohmic contact with Au electrodes, while CuO and ZnO have formed Schottky (non-ohmic) contact with Au electrodes. Since both sides of the same metal electrodes are used, it creates a double Schottky barrier. One of the two junctions is always reverse biased irrespective of the polarity of bias applied. The band bending and barrier heights are shown in Fig. 3 (k-m).

It is observed that having a heterojunction barrier, such as Schottky contacts, induces selectivity of specific gas in the oxide-based sensors. Therefore, these contacts may have contributed to the selective identification of gases.

Here, CuO and NiO are p-type semiconductors, while ZnO is an n-type semiconductor. The typical carrier type in these binary oxides arises because of particular defect chemistry[29]. The p-type oxides have metal vacancies whereas n-type oxides have oxygen vacancies as the dominant type of defect. These give rise to the acceptor and donor levels within the forbidden gap respectively. In this case, the thin film fabrication was done under significant oxygen partial pressures (30:10 SCCM of Ar and $O_2$ ratio). It ensures high lattice oxygen content in films, increasing metal vacancies for p-type and reducing oxygen vacancies for n-type. Therefore, the p-type films are more conducting than the n-type oxides under oxygen-rich deposition conditions.

Oxides, particularly ZnO conductivity (significantly just above room temperature), are strongly affected by the atmospheric oxygen content and moisture content. Some of our group's recent papers have reported protonic conductivity on ZnO nanoparticle surfaces, giving rise to the metal-like positive coefficient of temperature and its transition to semiconductor-like behavior at temperatures [29]. We also explored the frequency dependence of the electrical conductivity of the same ZnO nanoparticles and its ambiance[30]. Nevertheless, such anomalous behaviors are a vital function of size, morphology, and surface defects controlled via processing conditions. The studies mentioned above were performed on ZnO “nanoparticles” of 20 nm size. They, more importantly, were prepared by wet chemical methods where there may be significantly different surface defects and, therefore, different adsorption dynamics.

The ZnO samples reported in this work are deposited by DC reactive sputtering, leading to 2D nanosheets like ZnO. For similar sputtered films, the resistance anomalous behavior is not registered[31]. Besides, as seen in the newly added data, ZnO forms a non-ohmic contact with Au electrodes, and the electrode interface dominated the electrical behavior and sensing due to a high barrier (0.72 eV). The Schottky barrier leads to the inherent electric field in the sensor material, and the same enhances the response of the material as the carriers in the space charge region are heavily depleted, and any small change in carrier density leads to a significant difference in the conductance (G) of the Schottky junction.

$$G = \frac{\eta k_B}{e\,A^*\,T} e^{\frac{e\phi}{k_B T}} \quad (3$$

Where, $\eta$ is the ideality factor of the junction. Thus, the band bending changes lead to the barrier height (ϕ) change and the enhanced response. However, the Schottky barrier often induces selectivity in response, which is difficult to achieve in oxide materials's intrinsic response[32, 33].

On the other hand, p-type oxides like NiO and CuO inherently prefer selective oxidation for some of the VOCs due to the solid catalytic properties of transition metals (oxides)[34, 35]. Besides, as discussed earlier, CuO also shows Schottky behavior with Au electrodes.

### 3.1.3. Gas Sensing Measurements and data curation

Many experiments were performed to generate the response dataset for the gas sensor array with response to selected gases. Here, sensor temperature, gas concentration, and gas type have been identified as primary parameters for the sensor output.

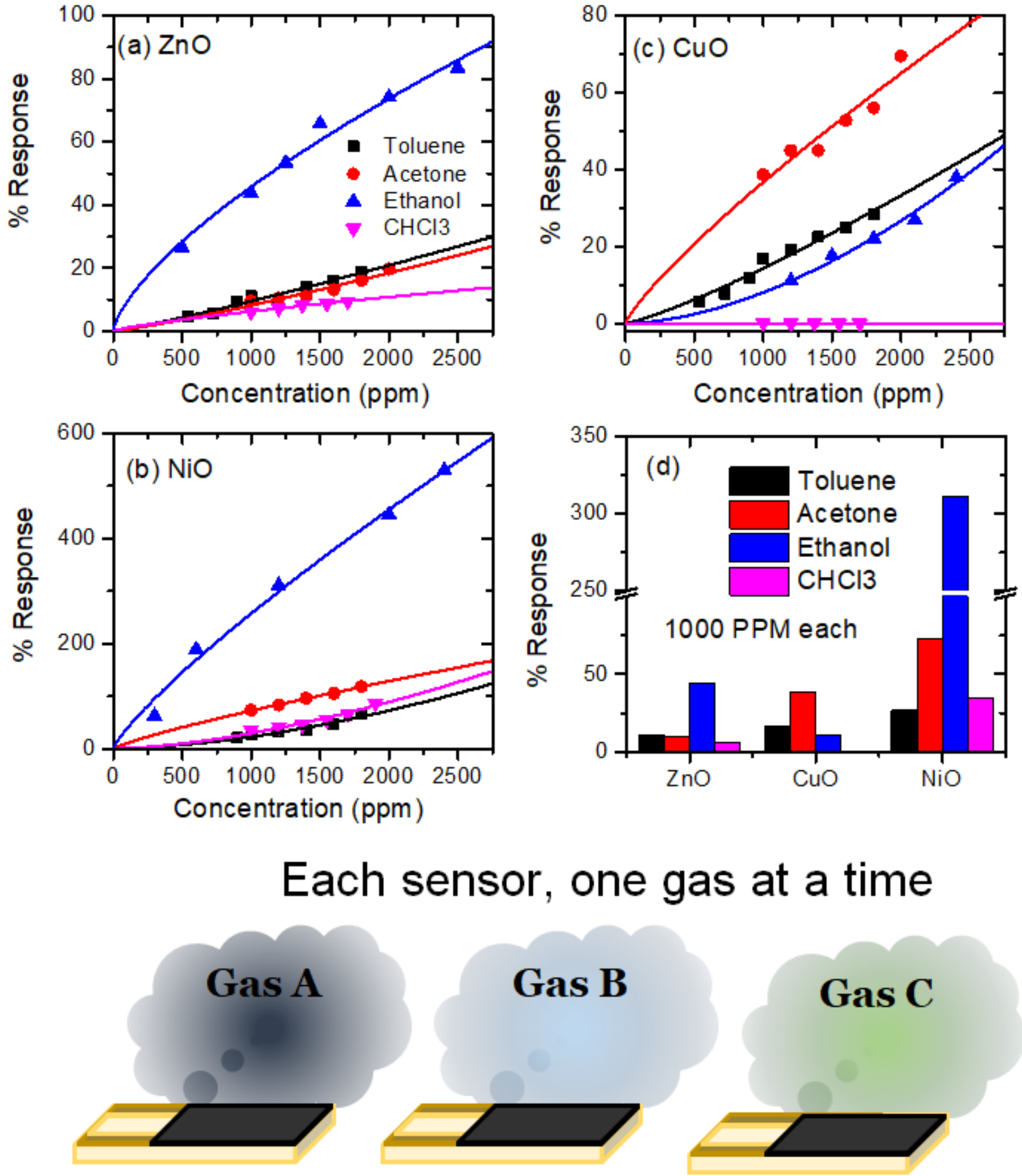

Fig. 4: The individual gas sensing results for four test gases, Toluene, ethanol, acetone, and chloroform, of the three samples (a) ZnO, (b) NiO, and (c) CuO at 200 °C. (d) The comparison of the response for 1000 ppm of each gas for each sensing electrode shows a preferred selectivity for ethanol in NiO and ZnO. In contrast, the CuO sensor does not offer any preferred selectivity. The solid lines denote power law fit. The bottom schematic shows the measurement protocol of same sensor being exposed to one gas at a time. at different concentrations. Overall, NiO showed a highly selective response to ethanol but a high response to all the gases. At the same time, ZnO had a consistently low response yet was selective to ethanol (See Fig. 4(d)). The actual data sets are shown in the Online Resource section ESM_7. The consistently high response NiO may be attributed to their commensurate (low and high) defect concentrations, respectively, as defects provide an active site for surface oxygen adsorption[29, 36].

The chemoresistive semiconductor gas sensors follow the power law behavior where the response can be predicted in any concentration range if another range response is available. The power law is given as shown in Eq. (4):

$$S = AC^{\beta} \tag{4}$$

Where S is the sensor response as denoted earlier, C is the concentration of gas (here in units of ppm) and is the exponent, which usually has a value between 0 and 1. Here, 1 denotes linear dependence, 0.5 denotes quadratic dependence, and 0 denotes independent (no response). As seen in Fig 4 (a-c), the measured response agrees very well with the power law behavior in all the cases; therefore, it can be easily extrapolated for low concentration range. Working with metal oxides offers the advantage of power law behavior[37, 38]. Therefore, extrapolating the response concentration curves down to lower parts per million (ppm) concentrations is possible without any loss of generality.

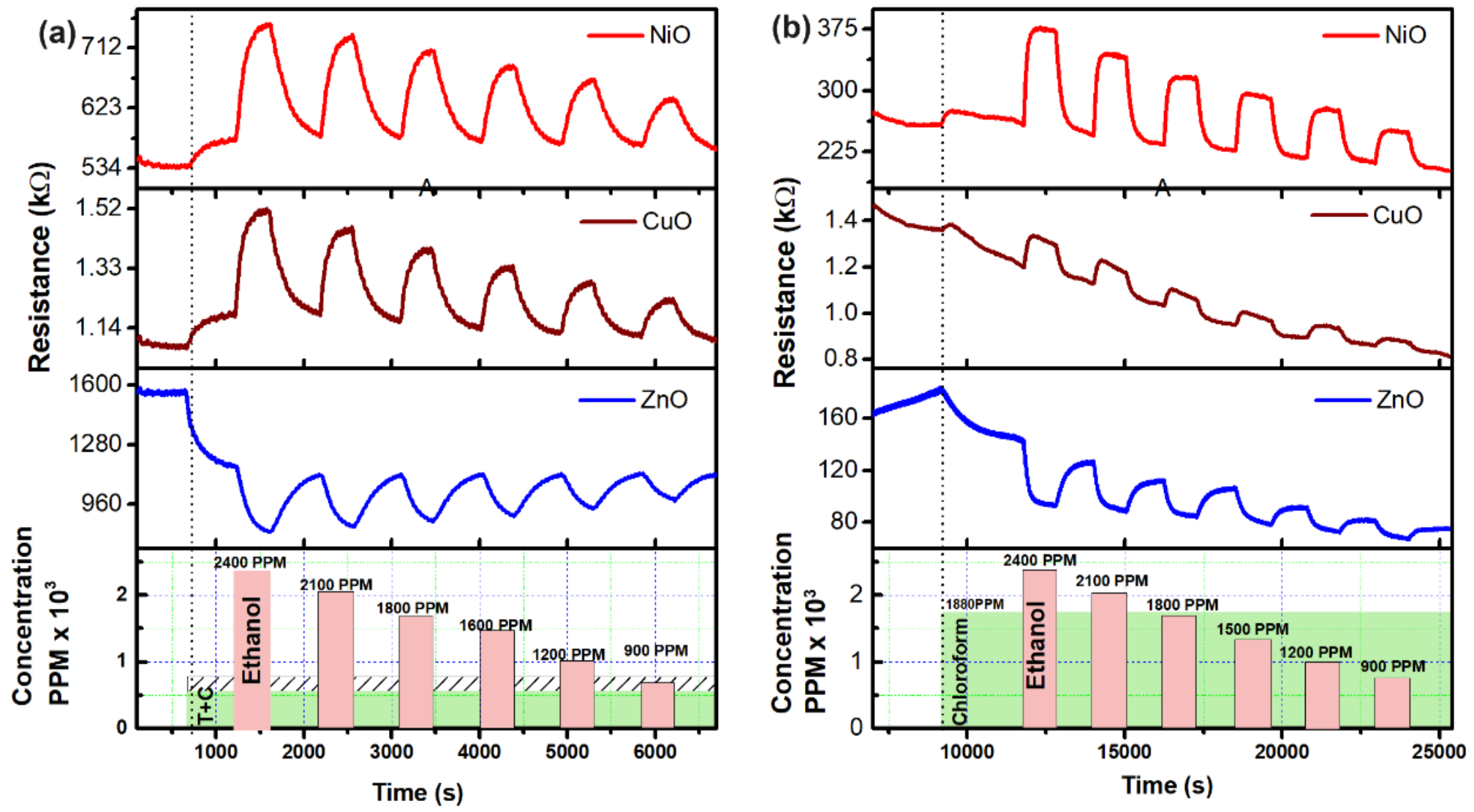

Fig. 5: The dynamic response in resistance data of ethanol sensing in presence of (a) chloroform, (b) chloroform and toluene as well as their computed response values for NiO in the absence & presence of a single interfering gas and a double interfering gas, respectively at 200°C. For other sensors the data is shown in Online Resource.

The single gas experiment results shown in ESM_7 in the Online Resource file are straightforward and are similar to how traditional gas sensors are reported. However, as mentioned earlier, detecting test gases becomes challenging in the presence of other potentially interfering gases. The experiments were designed such that a predetermined concentration of the interfering species is first supplied as a background flow in the chamber, followed by the introduction of the test gas (2-gases) to assess the impact of the interfering species (other gas) on the primary analyte (test gas ethanol). Calculations were made using the response values after varying the test gas concentration. The two interfering gases were maintained constant in the next series of trials (3-gases) while the test gas concentration was altered. The representative data for ethanol response in chloroform (2-gases) and in Toluene + chloroform (3-gases) have been shown in Fig. 5(a and b). The Online Resource section displays the other data sets in ESM_8 and S9 for two and three gases, respectively.

The values of response calculated here for 2-gases and 3-gases depict that the presence of any other VOCs led to a drastic reduction in response, as seen in Fig. 6. The response values calculated using eq (1) data for response in the absence and presence of a single interfering gas and a double interfering gas is shown in Fig. 6 for all the three sensors. Similar results are obtained when the treatment is done for all sensors and/or permutation – combinations of the gases.

Therefore, the mixture of gases (shown by red, black, and green lines in Fig. 6) is a substantially different condition from that of the single gas exposure (indicated by blue lines). The effect is more pronounced in the case of NiO, as seen in Fig. 6(a) and (b). Here, we found that the sensor does not obey the power law depicted in Eq (4) when exposed to such a mixture. However, the same may be modified to incorporate the shift in response leading to change in the response by an arbitrary value α, such that

$$S = \alpha + AC^{\beta} \quad (5)$$

This equation has been found to fit better, as seen in Fig 6. In some cases, the value of α is so significant that it affects the limits of detection substantially.

The following discrepancies form the basis of our study. Although the use of these "classical "oxide systems has been clichéd, the practical aspects of application still suffer from challenges such as

1. Lack of selectivity,
2. Practically extended response (and recovery times) to saturation (and recovery)
3. Drift in the sensor baselines

These often lead to inconsistencies; practically, there are no ways to avoid them as these are materials' intrinsic properties to a large extent. One may optimize the design to minimize them but need help eliminating them. Therefore, our research focuses on obtaining a natural response within a consistent timeframe, specifically assessing cross-reactivity within complex mixtures. We considered the real-world environmental factors, such as surface non-recovery and surface passivation, which become particularly relevant once the sensor reaches saturation, potentially leading to a loss of its inherent activity.

Increasing the operating temperature could have improved some aspects like response and recovery times, enhanced sensitivity, or better baseline. See the results obtained by the same sensors at 350°C shown in ESM_10 in the Online Resource file. Not only the energetic cost of continuously operating sensors from 200 to 350°C is substantially significant, but constant exposure to very high temperatures leads to several other issues.

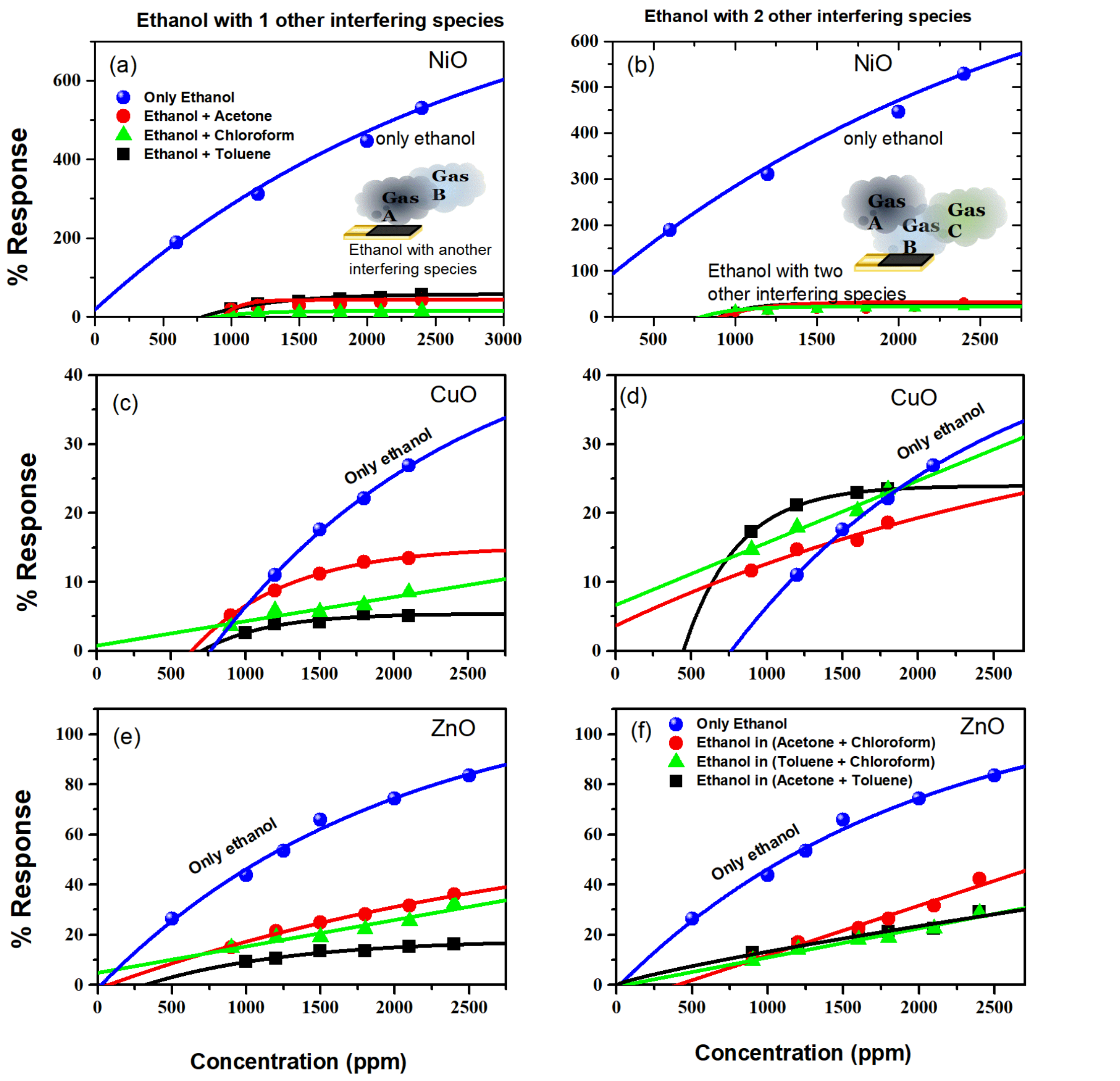

Fig 6. The response to single gas present at a time alongside the same gas present in (a, c, e) one or (b, d, f) two more gases simultaneously investigated for all possible combinations. The top row is for NiO, second is for CuO and third is for ZnO.

Our prime focus was to address the challenges of using metal oxide sensors regarding properties, including drift, high recovery, and response time in complex mixtures. To tackle these issues effectively, we successfully demonstrated the utility of our algorithms in machine learning techniques. Our approach involves various preprocessing steps, including data normalization, smoothening, and detecting outliers through our machine-learning algorithm. These techniques collectively address baseline variations and other potential problems, enabling us to predict and manage variations and drift situations more accurately and reliably.

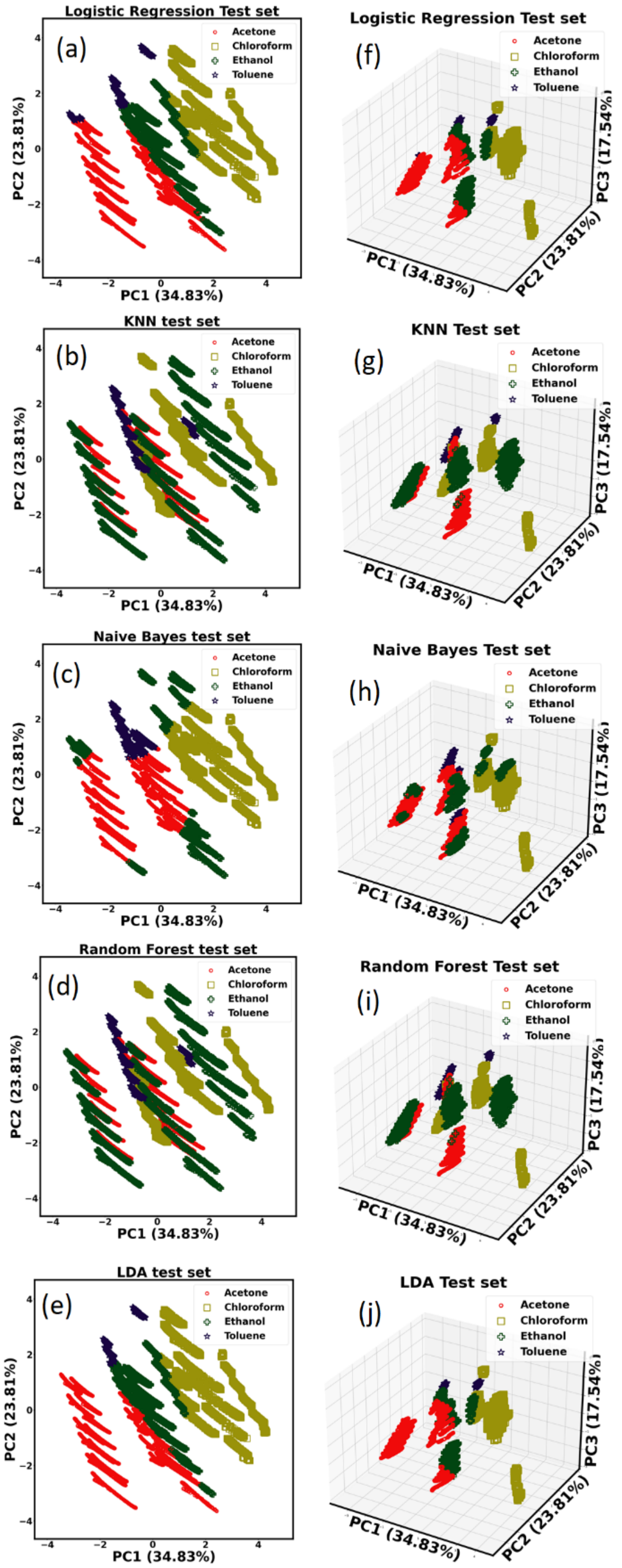

Fig. 7: 3-gases dataset: (a-e)2D and (f-j) 3D classification plots in 1st and 2nd columns respectively.

## 3.2. Classification and regression analysis of gases using machine learning

### 3.2.1. Gas classification

To reduce the complexity of the data while preserving trends and patterns, we used Principal Component Analysis (PCA)[39] on the sensor signal response. The variances of first 5 principal components (PC1, PC2, PC3, PC4, and PC5) are shown in Table S3 for 1-gas, 2-gases and 3-gases datasets. A pictorial representation of the variability of the first 5 PCs has been shown in ESM_11.

Here, we formulated the task as a classification problem to classify the gaseous chemicals, i.e., Acetone, Toluene, Chloroform, and Ethanol. The classification models were developed using some supervised learning techniques, e.g., Logistic Regression[40], K-Nearest Neighbor (KNN)[41], Naïve Bayes (NB)[42], Random Forest (RF)[43] and Linear Discriminant Analysis (LDA)[44], based on the PCA results for the gas classification. Different plotted points were dispersed depending on the type of chemicals used, as shown in Fig. 7 and Fig.s S12-S13. By considering PC1 and PC2, we obtained the 2D plots of Fig. 7 and Fig.s S12-S13 over three datasets. In this instance, PC1, PC2, and PC3 were also employed to produce 3D graphs. Here, we can visualize the qualitative performances of the employed ML models. For example, in Fig. 7 (b), (d), we can observe that KNN and RF have identified Ethanol samples correctly. However, as seen in Fig. 7 (a), (c), (e), many Ethanol samples have been misclassified as Acetone by logistic regression, NB, and LDA. In this instance, PC1, PC2, and PC3 were also employed to produce 3D graphs. In logistic regression[40], the training procedure employed the one-vs-rest scheme since our task involves multiple classes. We used cross-entropy loss and L2 regularization here[45]. In KNN[41], empirically the number of nearest neighbors was set to five, and the distance metric was chosen as Euclidean. In NB[42], every pair of features is conditionally independent given the class variable value, which is a supervised learning technique based on Bayes' theorem. To classify our data, we employed the Gaussian Naïve Bayes method. The RF and Extra-Trees methods are two averaging algorithms based on randomized decision trees that we employed[43]. Each algorithm uses a perturb & combine method that is tailored for trees. It means adding randomization to the classifier design results in creating a diverse group of classifiers. The average forecast of the individual classifiers is used to represent the ensemble prediction. Using Bayes' rule and fitting conditional class densities to the data, LDA[44] produces a linear decision boundary for classification. The model assumes that all classes have the same covariance matrix and fit a Gaussian density to each class. Fig.s S11-S12 and Fig. 5 display the 2D and 3D plots of the three datasets obtained after classification using the above-employed methods.

In Table 1, we present the accuracies obtained by the employed models. Here, KNN and random forest attained good accuracies for all three datasets, in contrast to the ML models like logistic regression, NB, and LDA. For 1-gas and 2-gases datasets KNN performed the best, and random forest attained the best result for the 2-gases dataset instead of their akin performances.

Table 1: Model performances over various gas mixture datasets

| | **Dataset \ Model** | **Logistic Regression** | **KNN** | **Naïve Bayes** | **Random Forest** | **LDA** |
|---|---|---|---|---|---|---|
| **Accuracy (%)** | **1-gas** | 65.56543 | **99.99802** | 71.75669 | 99.99679 | 60.06396 |
| | **2-gases** | 42.21952 | 99.81154 | 60.93418 | **99.82108** | 39.78625 |
| | **3-gases** | 38.73490 | **99.03290** | 51.83471 | 98.70436 | 39.76216 |

In Fig.s S12-S13 and Fig. 7, we can also comprehend misclassification results produced by logistic regression, NB, and RF. For example, in Fig. 7 bottom-left, it can be seen that the Ethanol part has been misclassified as Acetone.

### 3.2.2. Regression analysis: quantification of gases in different mixtures

In this analysis, we found that the KNN-based regression[46] significantly exceeded the other algorithms in terms of performance when compared with some other contemporary models, such as Artificial Neural Network (ANN), RF, Decision Tree, and Linear Regression[43, 45-48]. The performance of the KNN relies on various parameters, such as the distance metric used to evaluate similar data points, the number of neighbors taken into consideration, and the weighting method used to aggregate their values. In this study, we attempted to enhance the effectiveness of the KNN in estimating the gas concentration in mixes. To decrease MSE and increase the $R^2$, which gauges how much variance can be explained by the model, we set out to identify the optimal set of parameters.

To fine-tune the model, we experimented with various distance metrics, such as Euclidean, Manhattan, and Minkowski, with p=3 and p=4[16]. We used two weighting schemes: distance and uniform, wherein closer neighbors have a higher weight, and we adjusted the number of neighbors taken into consideration, ranging from 1 to 10. The model's performance was checked by applying cross-validation on the training and validation sets, and the optimum set of parameters was decided based on the parameters with the lowest MSE and optimum $R^2$. During the hyperparameter tuning procedure for the KNN regression, the best parameter choices for each gas mixture were identified. For all the datasets, i.e., 1-gas, 2-gases, 3-gases, the Euclidean distance metric, the five nearest neighbors, and distance weighting were the most efficient choices. Encouraging results were obtained while analyzing the algorithm's performance with all these ideal parameter configurations.

Table 2: Prediction performance of KNN regression on 1-gas, 2-gases, and 3-gases datasets.

| Dataset | Gas Name | RMSE | MSE | MAE | NRMSE | $R^2$ | LoD | LoQ |
|---|---|---|---|---|---|---|---|---|
| 1-gas | Acetone | 0.00086 | $7.43\times10^{-7}$ | 0.00001 | 0.00114 | 0.99997 | 0.00344 | 0.01146 |
| | Toluene | 0.00082 | $6.77\times10^{-7}$ | 0.00001 | 0.00109 | 0.99997 | 0.00328 | 0.01095 |
| | Ethanol | 0.00076 | $5.82\times10^{-7}$ | 0.00001 | 0.00101 | 0.99997 | 0.00304 | 0.01015 |
| | Chloroform | 0.00153 | $2.35\times10^{-6}$ | 0.00004 | 0.00203 | 0.99990 | 0.00611 | 0.02039 |
| 2-gases | Acetone | 0.00131 | $1.72\times10^{-6}$ | 0.00002 | 0.00319 | 0.99996 | 0.00957 | 0.03190 |
| | Toluene | 0.00094 | $8.98\times10^{-7}$ | 0.00001 | 0.00226 | 0.99998 | 0.00678 | 0.02260 |
| | Ethanol | 0.00095 | $9.21\times10^{-7}$ | 0.00001 | 0.00230 | 0.99998 | 0.00692 | 0.02309 |
| | Chloroform | 0.00194 | $3.79\times10^{-6}$ | 0.00006 | 0.00466 | 0.99992 | 0.01400 | 0.04669 |
| 3-gases | Acetone | 0.00163 | $2.67\times10^{-6}$ | 0.00005 | 0.00393 | 0.99994 | 0.01179 | 0.03932 |
| | Toluene | 0.00204 | $4.19\times10^{-6}$ | 0.00006 | 0.00496 | 0.99991 | 0.01488 | 0.04961 |
| | Ethanol | 0.00196 | $3.87\times10^{-6}$ | 0.00005 | 0.00474 | 0.99992 | 0.01422 | 0.04742 |
| | Chloroform | 0.00342 | $1.17\times10^{-5}$ | 0.00020 | 0.00825 | 0.99976 | 0.02478 | 0.08260 |

Table 2 presents the prediction performances of KNN regression on 1-gas, 2-gases, and 3-gases datasets, respectively, regarding RMSE, MSE, MAE, NRMSE, R2, LoD, and LoQ. The model successfully predicted the target variable for the 1-gas mixture with $R^2$ of more than 0.99, showing its high prediction performance. Also, it was determined that the corresponding errors (RMSE, MSE, MAE, and NRMSE) were shallow. The model also obtained an outstanding $R^2$, i.e., greater than 0.99 for the 2-gases and 3-gases mixtures, implying a solid connection between observed and predicted values. Also, errors were near zeros, implying comparatively smaller magnitudes of the prediction mistakes. The model also excelled in other performance metrics, e.g., LoD and LoQ, when examined in the instances of the 1-gas, 2-gases, and 3-gases datasets. In Fig. 8, we present the regression plots obtained using KNN regression, where the x and y axis denote expected and obtained chemical concentrations separately for Acetone, Toluene, Ethanol, and Chloroform over 1-gas, 2-gases, and 3-gases datasets. As mentioned earlier, we have used ANN, Random Forest, Decision Tree, and Linear Regression for comparative prediction analysis. The ANN can learn and adapt to new data, making it a powerful tool for solving complex problems. However, ANN requires a lot of data and computational power to train and optimize, and its results may only sometimes be interpretable[45]. Here, in the ANN model, we had one neuron on the output layer that matched the concentration of the varying gas. The model comprised six hidden layers containing 128, 256, 512, 64, and 32 neurons. All hidden layers employed the ReLU (Rectified Linear Unit) activation function to capture the non-linearity[49]. We utilized a linear activation function in the output layer. The learning parameters for the ANN model were optimized on the training set using the Adam optimization function. Here, the training effectiveness was assessed using the loss function MSE. The following hyper-parameters were empirically fixed on the validation set: learning rate = $10^{-3}$, Adam's first and second moment estimates 0.9 and 0.999, and zero-denominator remover = $10^{-7}$.

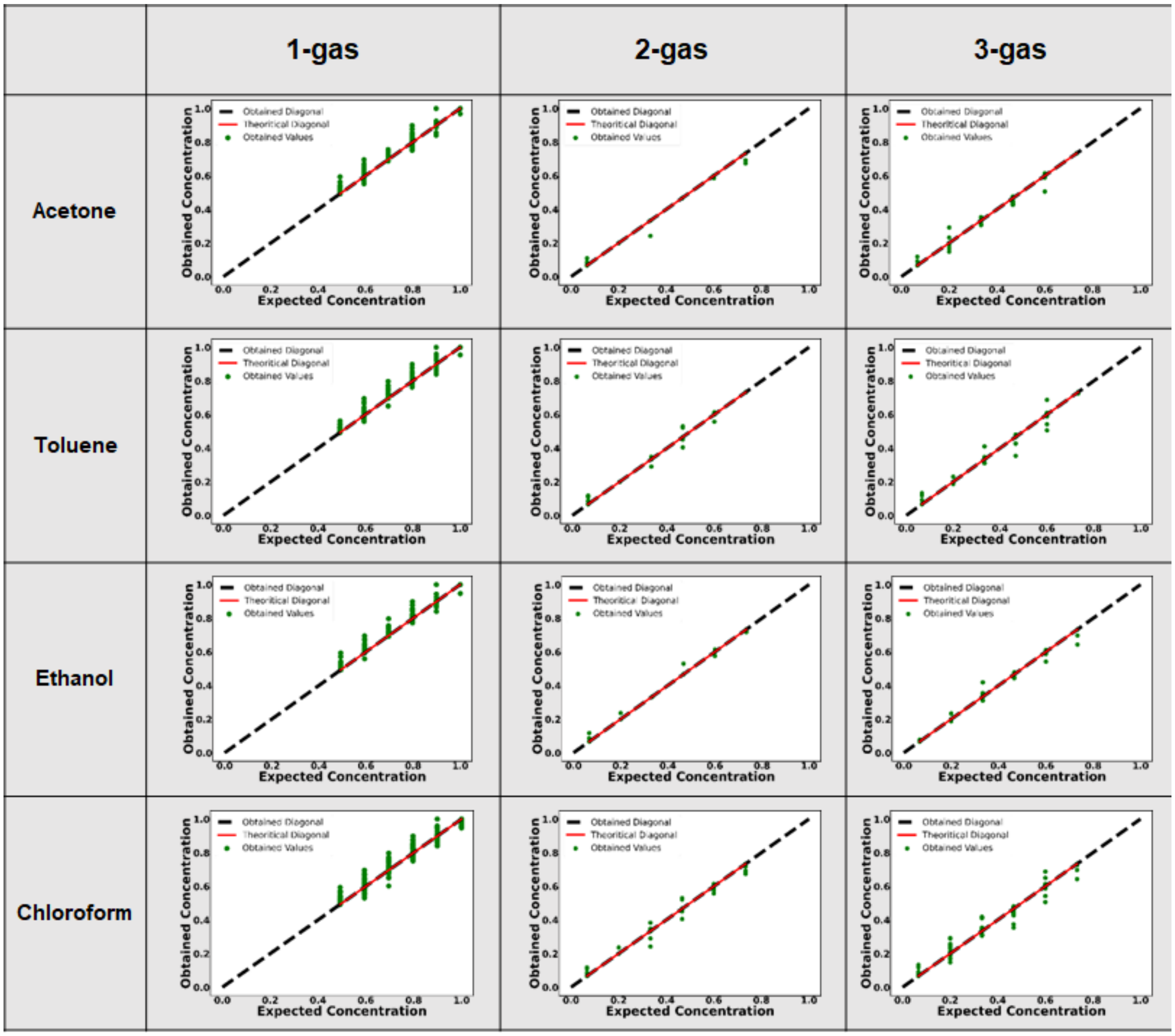

Fig. 8: Prediction plots of KNN regression: 1st Column: 1-gas dataset, 2nd Column: 2-gases dataset and 3rd Column: 3-gases dataset. Row-wise, the prediction of Acetone, Toluene, Ethanol, and Chloroform, respectively.

In linear regression[47], we model the relationship between the dependent and one or more independent variables. Here, we identify the line of best fit that minimizes the sum of squared errors between the predicted and actual values. In decision tree regression[48], we use a tree-like model of decisions and their possible consequences for prediction. However, they can be prone to overfitting and may need to be more accurate in certain situations. Random forest[43] ensembles multiple decision trees to improve performance and reduce overfitting. It randomly selects a subset of features and data samples for each tree to make it robust to noise and outliers. It also offers feature importance ranking and can handle missing data. However, it may perform poorly on imbalanced datasets and can be computationally expensive for large datasets.

In Tables S4, S5, and S6, we compare the experimental results obtained on 1-gas, 2-gases, and 3-gases datasets using KNN regression, ANN, random forest, decision tree, and linear regression models. The evaluation results regarding metrics RMSE, MSE, MAE, NRMSE, $R^2$, LoD, and LoQ are shown here for predicting Acetone, Toluene, Ethanol, and Chloroform gases. Overall, it can be observed from these tables that KNN regression outperformed here over all the datasets.

For better visibility, we summarize Tables S4, S5, and S6 and compare the results concerning only $R^2$ in Table S7. The KNN-based regression technique achieved exceptional performance across all three datasets, achieving $R^2$ of more than 0.99, in stark contrast to the contemporary regression models, such as ANN, random forest, decision tree, and linear regression. Only in the 2-gases dataset, for chloroform prediction, random forest performed slightly better than KNN regression. The performance of the random forest was also quite similar to the KNN regression here.

# 4. Discussion

Although metal oxide thin films are the most successful sensor materials, the major limitation of these materials is their lack of selectivity. The traditional way of characterizing gas sensor devices involves one-by-one exposure to each gas and characterizing the sensitivity as shown in Fig. 4. In such cases, the sensor may show a significantly preferred sensitivity, called selectivity towards a particular gas (like ZnO and NiO shows for ethanol and CuO shows for acetone in the 1-gas case presented here). However, it gets challenging when another potential interfering gas exists in the atmosphere. Although the other interfering gas may not have high sensitivity when present individually, it adversely affects the response towards other gases through interference. For instance, when more than one gas is present the response drastically reduces as seen in Fig. 6. Thus, ethanol gas response when studied in the presence of other single or double gases, the response is substantially reduced (sometimes by order of magnitude as seen in NiO case).

Therefore, using conventional analysis methods, gas mixtures are challenging to analyze using a single sensor or even an array of sensors. Albeit, the sensors utilized in the study are robust, and sensitive and show good microstructural traits as required for an ideal metal oxide material for high responsivity[50, 51].

Subsequently, we employed ML-based methods to analyze the sensor array response of such a complex mixture where there is maximum cross-reactivity for one sensor (CuO) while the other two show some preferred selectivity (NiO and ZnO) towards ethanol. Our analysis involved ML algorithms like RF, KNN, Decision Tree, Linear Regression, Logistic Regression, Naive Bayes, LDA, ANN, and SVM which are used to find the patterns in response. Among these, RF and KNN gave the best results with extraordinary accuracy of more than 99%. The algorithms could classify and identify the gas type and reasonably estimate the gas concentration of the varying chemicals for 1-gas, 2-gases, and 3-gases datasets.

Table 3: Comparative analysis with some state-of-the-art studies.

| **No. of sensors** | **No. of gases together** | **Complexity** | **Models Used** | **Ref.** |
|---|---|---|---|---|
| **1** ($WO_3$) | **3** (CO, $O_3$, $NO_2$) | Medium | SVM | [52] |
| **4** (Commercial MOS Sensors TGS 2600, TGS2602, TGS 2610, TGS 2620) | **2** ($NO_2$, CO) | Medium | BPNN + CNN | [53] |
| **1** (Graphene) | **36** VOC Receptors | High | PCA + RF | [54] |
| **1** ($SnO_2$ Nanowires) | **5** (Acetone, Ammonia, $H_2$, $H_2S$, Ethanol) | Medium | SVM | [55] |
| **1** (Graphene) | **2** ($NH_3$, $PH_3$) | Low | PCA + LDA | [56] |
| **1** (ZnO) | **3** *(separate)* ($H_2S$, $NH_3$, $CO_2$) | Low | NB + LR + SVM + RF | [57] |
| **1** ($SnO_2$) | **4***(separate)* (Formaldehyde, Methanol, Propanol, Toluene) | Low | FFT + DWT (SVM + RF + MLP) | [58] |
| **5** (Commercial MOS Sensors TGS2600, TGS2610, TGS2611, TGS2602, TGS2620) | **2** ($CH_4$, CO) | Medium | PCA + ICA + KPCA + KNN + MVRVM | [59] |
| **1** (ZnO) | **7** (separate) (Toluene, Acetone, $NH_3$, Ethanol, 2-Propanol, Formaldehyde, Methanol) | Low | PCA | [60] |
| **3** ($SnO_2$, Au/ $SnO_2$, AuPd/ $SnO_2$) | **2** (Methane, Propane) | Medium | SSA | [61] |
| **3** (CuO, ZnO, CuO-ZnO) | **4** (Methanol, Acetonitrile, Isopropanol, Toluene) | Low | t-SNE + SVM | [62] |
| **3** (ZnO, NiO, CuO) | **4** (Ethanol, Acetone, Toluene, Chloroform) | Medium | KNN + ANN + RF +DT + LiR + LR +NB + LDA | This work |

**SVM:** Support Vector Machine, **BPNN:** Back Propagation Neural Network,
**CNN:** Convolutional Neural Network, **PCA:** Principal Component Analysis,
**RF:** Random Forest, **LDA:** Linear Discriminant Analysis,
**NB:** Naïve Bayes, **LR:** Logistic Regression,
**FFT:** Fast Fourier Transform, **DWT:** Discrete Wavelet Transform,
**MLP:** MultiLayer Perceptron, **ICA:** Independent Component Analysis,
**KPCA:** Kernel Principal Component Analysis, **KNN:** K-Nearest Neighbors,
**MVRVM:** Multivariate Relevance Vector Machine, **SSA:** Statistical Shape Analysis,
**t-SNE:** t-Distributed Stochastic Neighbor Embedding, **DT:** Decision Tree,
**LiR:** Linear Regression, **ANN:** Artificial Neural Network.

The level of complexity of data and the resources used, such as no of sensors in the array, no gases studied, the model used, and the complexity of data in this study have been compared with that of other studies reported in the literature and presented in Table 3. For instance, Djedidi O. *et al.*[52] created a method to use a single temperature-modulated MOS sensor and a data-driven model to detect and identify various gas species and their mixtures. By taking the characteristics from dynamic curves and introducing a four-sensor array, Chu J. *et al.*[53] could distinguish between 11 different $NO_2$ and CO mixes and identify different target gases using BPNN. The categorization of VOC species and concentrations using a 108-device graphene-based sensor array swept at high speeds has been shown in the study conducted by Capman N S S. *et al.*[54]. To increase selectivity, the array was functionalized with 36 different chemical receptors. All devices were virtually probed simultaneously to gather a cross-reactive data set for ML algorithms. To discriminate between 5 distinct reducing gases, two multi-sensor chips made of $SnO_2$ nanowires covered with Ag and Pt NPs were combined by Thai N X. *et al.* [55]. The "brain" of the system (based on the SVM) is trained using a first dataset of 4D points, and the sensor performance is tested using any subsequent point. With practical machine learning algorithms and MDS (Molecular Dynamic Simulations), Huang S. *et al.*[56] have shown an ultrasensitive, highly discriminative graphene nanosensing platform for detecting and identifying NH3 and PH3 at room temperature. Kanaparthi. *et al.*[57] have developed an analytical technique that uses a single chemiresistive ZnO gas sensor to detect $NH_3$, $CO_2$, and $H_2S$ gases selectively at significantly low power consumption. To anticipate the gas present in the air, ML techniques including NB, LR, SVM, and RF were used for the data comprised of sensor responses and ternary logic. Over a single chemiresistive sensor, Acharya S. and coworkers[58] used signal transform methods combined with ML technologies, which allowed for accurate quantification and selective identification of the tested VOCs. The feature extraction technique suggested in the study by Xu Y. *et al.*[59] is based on KPCA. Qualitative identification of mixed gas is made possible by the binary mixed gas identification model of the KNN classification method. A regression approach based on MVRVM was suggested to obtain quantitative gas concentration detection for the qualitative identification findings. Sett A. *et al.*[60] used ZnO nanorods to create a susceptible, stable, and reliable VOC sensor. In reaction to three VOCs, the sensor showed high responsiveness and stability. Features were taken out and supplied into PCA as input. Ref[61] shows that applying statistical shape space pre-processing to the signal of temperature-modulated metal oxide gas sensors improves the selectivity of gas identification with an ANN-based ML algorithm compared to other signal processing methods like PCA, DWT, polynomial curve fitting, and data normalization. Intrinsic CuO and ZnO heterostructures with different weight percentages of CuO–ZnO were made and used as resistance sensors to find four volatile organic compounds. The SVM algorithm with stacked k-fold cross-validation was used for classification and measurement, the MLR method was used[62].

On the other hand, in this work, we have used only three sensors that operate at the same temperature and show a distinct mix of selective (NiO and ZnO) and non-selective sensors (CuO) for ethanol vapors. Using two algorithms we obtained the best possible classification (qualitative) and regression (quantitative) identification of gases. Moreover, the gases identified in the study are highly likely to indicate underlying physiological conditions in several diseases like diabetes, lung cancer, heart diseases, etc, Therefore, sensor and analysis studies have high significance for biomedical diagnostics and point-of-care devices. In Table 3. it may be seen that the current study demonstrates excellent recognition capabilities with minimal elements in the sensor array.

# 5. Conclusion

In this study, we fabricated a gas sensor array consisting of three metal oxides, i.e., ZnO, NiO, and CuO. NiO showed ohmic contact with Au, while others showed Schottky. Each sensor in the array was extensively characterized using state-of-the-art surface and material characterization techniques (e.g., SEM and XRD). Each of these materials is highly responsive to a large number of gases, generating cross-reactive and complex chemiresistive signals, it can be used to detect many gases. Moreover, it is observed that when more than 2 VOCs are present in the atmosphere, the sensor's response is drastically different. ML algorithms have been used to classify and predict the levels of individual gases in mixtures to handle such complex data sets. To get the best algorithms out of several that we tried, the parameters of the algorithms have been extensively optimized toward the classification and prediction of different analyte gases. We anticipate that the proposed sensor array can be used for the analysis of different VOCs in complex mixtures (e.g., breath) for non-invasive diagnostic of disease and its monitoring at the point-of-care. The developed sensor array could be used to diagnose different diseases at the point-of-need non-invasively, which can improve the quality of life of individuals.

Although it was not explicitly mentioned, the data used in the study has been curated on a long duration of several months (6-8 months), thus the long-term stability is evident and degradation with time if any, is also integrated in the machine learning analysis. However, there could be a stand-alone study on the stability aspects.

**Statements and Declarations**

**Competing Interests**

The authors have no competing interests to declare that are relevant to the content of this article.

**Funding**

The authors are thankful to the SERB core research grant (CRG/2022/006973) Govt. of India for the funding support received. The Central Instrumentation Facility of IISER Thiruvananthapuram is also acknowledged for the XRD and SEM facilities.

**Data Availability**

The data is available with the corresponding author upon reasonable request.

**Online Resource**

The Supplementary Material includes the sample fabrication details, the device design, the I-V characteristics, The Gas sensing data as well as the different ML model parameters, etc.

## Online Resource

# Metal Oxide-based Gas Sensor Array for the VOCs Analysis in Complex Mixtures using Machine Learning

**Shivam Singh[1], Sajana S[1], Poornima Varma[2], Gajje Sreelekha[3], Chandranath Adak[3,*], Rajendra P. Shukla[4,*], Vinayak B. Kamble[1,*]**

[1]School of Physics, Indian Institute of Science Education and Research Thiruvananthapuram, 695551 India.

[2]Dept. of CSE, Indian Institute of Information Technology Lucknow, Uttar Pradesh 226002, India.

[3]Dept. of CSE, Indian Institute of Technology Patna, Bihar 801106, India.

[4]BIOS Lab-on-a-Chip Group, MESA+ Institute for Nanotechnology, Max Planck Center for Complex Fluid Dynamics, University of Twente, P.O. Box 217, 7500 AE Enschede, The Netherlands.

Corresponding authors: Chandranath Adak, Rajendra P. Shukla, Vinayak Kamble.

# 1. Various Volatile organic Compounds in exhaled breath as biomarkers for diagnostics and our approach

Exhaled Breath includes several volatile chemicals, most present in minimal ppb concentrations. Therefore, the proportion of exogenous VOC can be used as a gauge for human biology and physiological health.

**Ethanol:** Numerous potential uses, including exhaled condition monitoring for the detection of smaller doses of ethanol, have lately gained considerable attention. Breath ethanol levels in a healthy individual are typically below 380 parts per billion. Nevertheless, this might increase to 2300 ppb in cases of alcoholism and a history of fatty liver[1-4].

**Acetone:** person who has elevated breath acetone (T2DM > 1.71 ppm, T1DM > 2.19 ppm (Type 1 diabetes mellitus (T1DM), an asymptomatic disease, is caused by the body's antibodies attacking and killing the beta cells that produce insulin in the pancreas. However, in type 2 diabetes mellitus (T2DM), the body generates inadequate quantities of insulin or becomes resistant, making it difficult to maintain normal blood sugar levels. It often develops in adults and correlates with lifestyle factors, including obesity and inactivity) may well go approximately 21 ppm[1-3]), which would ring alarming bells of diabetes. A biochemical disorder like diabetes mellitus affects over 400 million people worldwide[4] linked to such occurrences.

**Toluene:** a high concentration of toluene in a person's breath has been recognized as a potential biomarker linked to lung diseases[5-7]. In addition, toluene has also been found to be a low-level VOC in type I and type II diabetes. This implies that toluene can also serve as a significant indicator or diagnostic tool for identifying symptoms associated with diabetes.

**Chloroform:** In contrast to the others, chloroform belongs to a broader class of compounds called trihalomethanes are linked to diseases like lung cancer[8, 9] thyroid abnormalities[10], etc.

Nevertheless, their concentration range is yet to be established. In humans, breathing indoor air or consuming substantial quantities of chloroform-containing liquids such as chlorinated water may cause fatal consequences[11, 12], and the same can be diagnosed by the presence of chloroform in breath[13].

In this study, we employed a gas sensor array based on targets that were sputtered with a Direct Current (DC) source to deposit the metal oxides CuO, NiO and ZnO. The responses of various analytes passing over it were recorded using four volatile compounds: ethanol, toluene, acetone, and chloroform. The initial phase was mixing a single analyte with synthetic air and recording the response (resistance vs. time) for each electrode at 200 °C. One electrode was utilized at a time. After that, an experiment was conducted using two gases simultaneously, with one analyte being kept constant at a specific concentration while the other was changed. There were twelve different potential combinations for the 200 °C parallel readings. Similarly, the reaction was measured while carefully purging three gases, with two maintaining constant and the third changing. The gas sensor array's electrical response was examined using ML techniques. We have used different ML algorithms to analyze the data from the Metal Oxide Semiconductor (MOS) sensor array and compared their performances for the simultaneous detection of four VOCs. The ML algorithm was used to perform two types of analysis: (i) *classification* to categorize the varying gas/ chemical and (ii) *regression* analysis to predict the concentration of the gas. Therefore, not only qualitative but quantitative detection of four VOCs simultaneously allows the detection of multiple diseases and monitoring of the health of individuals. The proof-of-concept demonstration using a sensor array combined with ML algorithms can potentially analyze individual VOCs in breath samples to provide diagnostic and therapeutic information in diseases outlined such as lung cancer, heart diseases, diabetes and fibrosis, etc. Further miniaturization and its application to point-of-care testing devices can improve diagnostics and treatment monitoring of diseases (e.g., cancer).

## 2. Thin film deposition details.

Table S1. Sputtering parameter and deposition conditions for all three oxides.

| **Deposition conditions of The oxide film** | **CuO** | **NiO** | **ZnO** |
|---|---|---|---|
| Base Pressure(mbar) | 9.80E-6 | 3.84E-6 | 7.80E-6 |
| Deposition pressure(mbar) | 2.42E-2 | 2.59E-2 | 5.10E-2 |
| Target (1 inch diameter) | Cu metal (3 mm thickness) | Ni foil (1 mm thickness) | Zn metal (3 mm thickness) |
| Argon flow rate (SCCM) | 30 | 30 | 30 |
| Oxygen flow rate (SCCM) | 10 | 10 | 10 |
| Substrate | Alumina + Glass | Alumina + Glass | Alumina + Glass |
| Voltage applied | 538 V | 343 V | 377 V |
| Current | 0.08 A | 0.26 A | 0.05 A |
| Deposition time (min) | 14 | 25 | 25 |
| Substrate temperature | RT | RT | RT |
| Rotation used | Yes | Yes | Yes |

Table S1 summarizes the DC magnetron deposition parameters of the sensor material thin film fabrication.

## 3. Gas sensor device Fabrication and dimensions

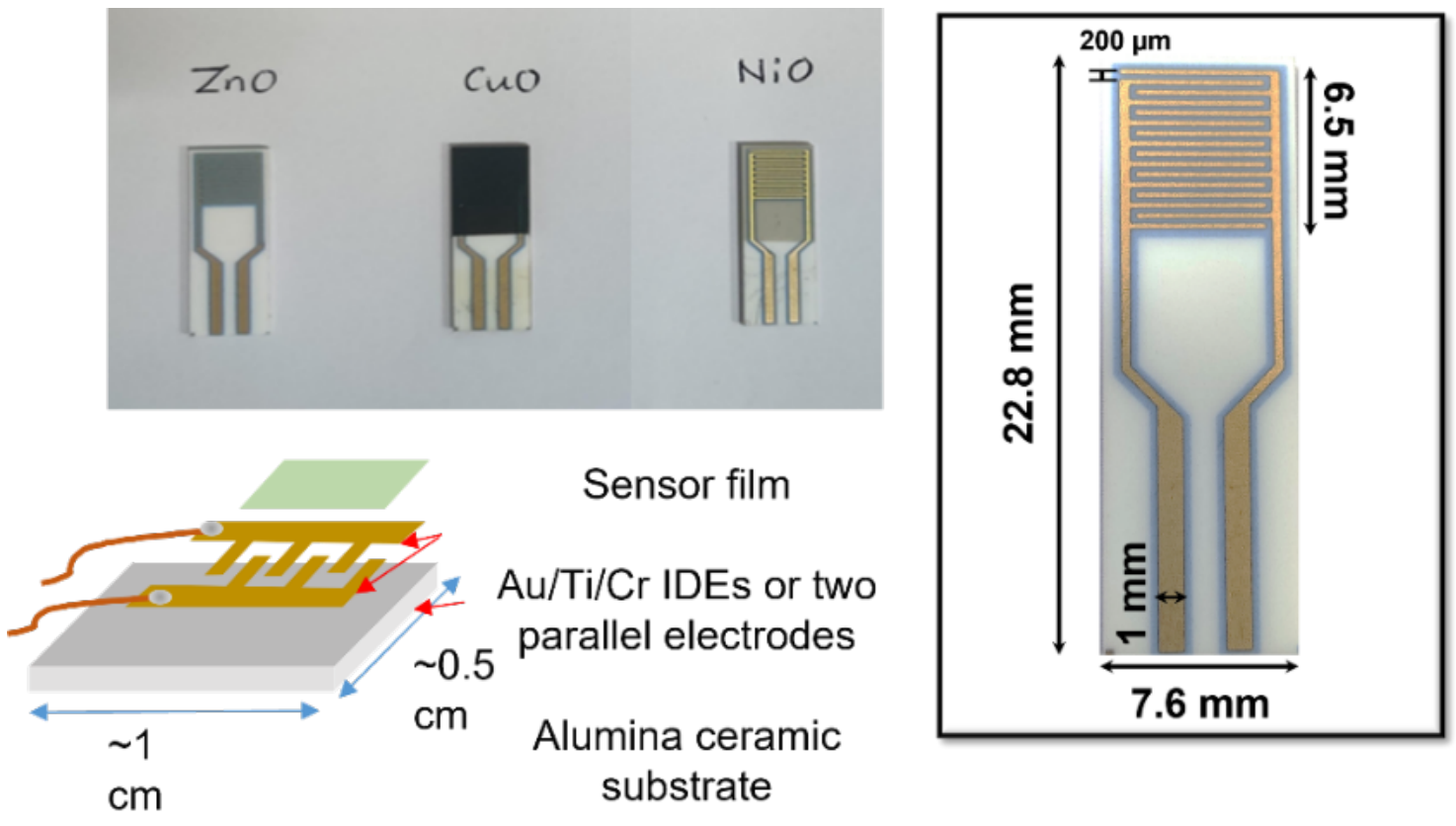

**ESM 1: The digital picture of the fabricated ZnO, CuO, and NiO sensors. The schematic diagram of the electrodes is shown in the bottom figure. The dimensions of the substrate and electrodes are shown in the adjacent figure.**

The electrode (Au) metal choice is crucial for the sensor response as a non-ohmic contact may be formed if the work functions alignment does not favor the bidirectional flow of current. A sensor design features an alumina substrate with interdigitated gold electrodes (IDEs). Here, to maintain the reproducibility of the nature of contact so that the data is consistent in repeated measurements, we chose prefabricated Au IDEs on the alumina substrates commercially available (by Metrohm India Pvt Ltd.).

The dimension is L 22.8 mm x W 7.6 mm x H 1 mm. The width and gap between the two electrodes are 200 µm.

## 4. Gas sensing system details and design.

The VOCs were generated using bubblers and vapor pressure data-based calculations were employed to estimate concentrations.

Synthetic air was bubbled over volatile organic compounds kept at a constant cooling of zero degrees Celsius by an MFC-controlled carrier gas flow to produce the test gas vapors. Since the vapor pressure of the analyte can be determined at a given temperature using the Antoine equation, it is possible to calculate the gas concentration by equilibrating it to the environment. Eq (S1) passes the attention;

To create the test gas vapors, synthetic air was bubbled over volatile organic compounds maintained at a constant cooling of zero degrees Celsius by an MFC-controlled carrier gas flow. Since gas concentration can be calculated by equilibrating the vapor pressure of the analyte to the atmosphere, as the vapor pressure of the VOCs is known at a given temperature. The concentration is given by eq;

$$C(ppm) = \frac{F_s}{F_s + F_c + F_D} \times 10^6 \quad \text{(S1)}$$

where, $F_d$ is the dilution gas flow rate in SCCM, $F_c$ is the input flow rate to the analyte in SCCM, and $F_s$ is the output flow rate of the vapor and can be calculated as shown in eq.

$$F_s = \left(\frac{P_{th}}{P_o - P_{th}}\right) F_c \quad \text{(S2)}$$

Here, $P_{th}$ is the thermodynamic vapor pressure of the analyte at that temperature, and $P_o$ is the atmospheric vapor pressure. More details can be found elsewhere[14, 15].

As mentioned in Section 2.1.4, the volatile liquids are used to generate the vapors of the desired gas for sensing. Here, the bubblers are maintained at a constant temperature, and the carrier gas is bubbled through the liquid in the thermostat to generate the vapors subjected to sensor exposure. Here, the concentrations of the vapors are mainly governed by constant temperature baths and the flow rate of the carrier gas to a certain extent. Therefore, the gas concentrations utilized were primarily governed by the generation rate and vapor pressure. The primary objective was to explore and investigate the region where the concentrations of these interfering biomarkers were high. Subsequently, the sensor response was recorded by introducing variable gas concentrations within this specific range, as mentioned above. This approach thoroughly examines and characterizes the sensor's behavior when exposed to various interfering gases at different concentrations. Besides, the gas sensing apparatus' practical limits, such as MFC resolution accuracy, primarily determined our study's interference gas concentration selection. We have been focusing on our system's capacity to manage intricate gas combinations while optimizing and miniaturizing them. Working with metal oxides has been a critical component of our strategy since it allows us to extrapolate response concentration curves to lower parts per million (ppm) concentrations. Tapping into the power of machine learning to improve our system's accuracy and forecasting powers at these lower concentrations is envisaged—the strategy to be rigorously trained and fine-tuned to produce highly accurate predictions even at sub-ppm levels.

## 5. Combinations of gases undertaken for sensor array response characteristics

Table S2: Combinations of VOCs used for 2-gases and 3-gases experiments.

| Variable VOC | 2-gases (Constant VOCs) | 3-gases (Constant VOCs) |
|---|---|---|
| **Ethanol** | Acetone (2000 ppm) | Chloroform (1000 ppm) + Toluene (720 ppm) |
| | Toluene (1880 ppm) | Acetone (600 ppm) + Chloroform (1000 ppm) |
| | Chloroform (1880 ppm) | Acetone (600 ppm) + Toluene (720 ppm) |
| **Acetone** | Toluene (1880 ppm) | Ethanol (900 ppm) + Toluene (540 ppm) |
| | Chloroform (1880 ppm) | Chloroform (1000 ppm) + Toluene (720 ppm) |
| | Ethanol (2400 ppm) | Chloroform (1000 ppm) + Ethanol (900 ppm) |
| **Toluene** | Chloroform (1880 ppm) | Acetone (1000 ppm) + Ethanol (900 ppm) |
| | Ethanol (1880ppm) | Ethanol (900 ppm) + Chloroform (1000 ppm) |
| | Acetone (2000 ppm) | Acetone (600 ppm) + Chloroform (1000 ppm) |
| **Chloroform** | Ethanol (2400 ppm) | Acetone (600 ppm) + Ethanol (900 ppm) |
| | Acetone (1600 ppm) | Ethanol (900 ppm) + Toluene (720 ppm) |
| | Toluene (1000 ppm) | Acetone (400 ppm) + Toluene (540 ppm) |

The various combinations of gases tested for generating the test data-set and training the algorithms have been described in Table S2.

6. **The correlation matrices of the various features for each of the gas mixture studied have been shown in ESM 2 below.**

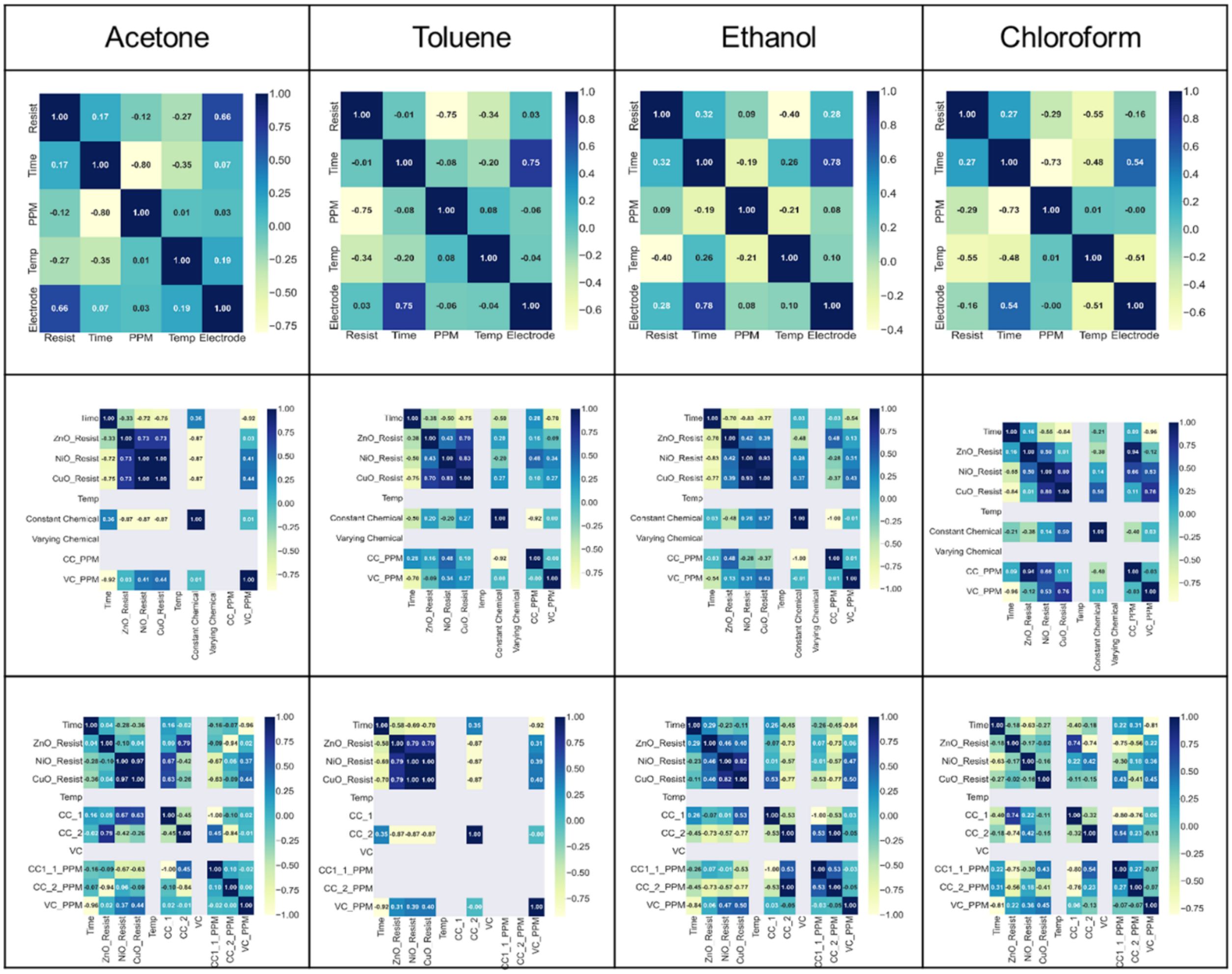

ESM 2: Correlation matrices: 1st row: 1-gas dataset, 2nd row: 2-gases dataset, 3rd row: 3-gases dataset.

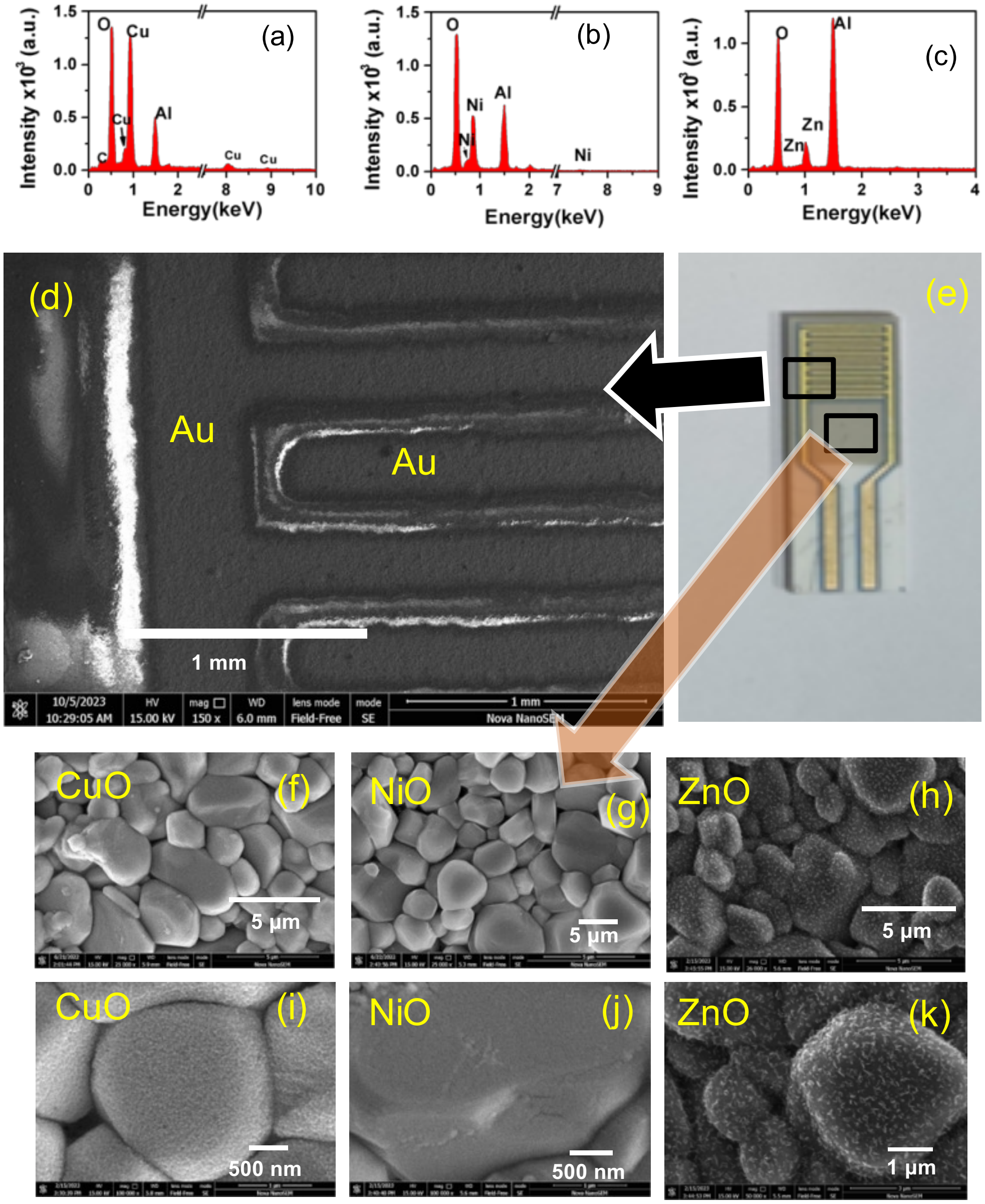

ESM 3: The Energy dispersive spectra showing compositions of (a) CuO, (b) NiO, and (c) ZnO thin films. (d) the SEM image of the Au interdigitated electrodes with oxide film deposited on top. (e) the digital photograph of the device to locate the region in the image (d) and (f-k).

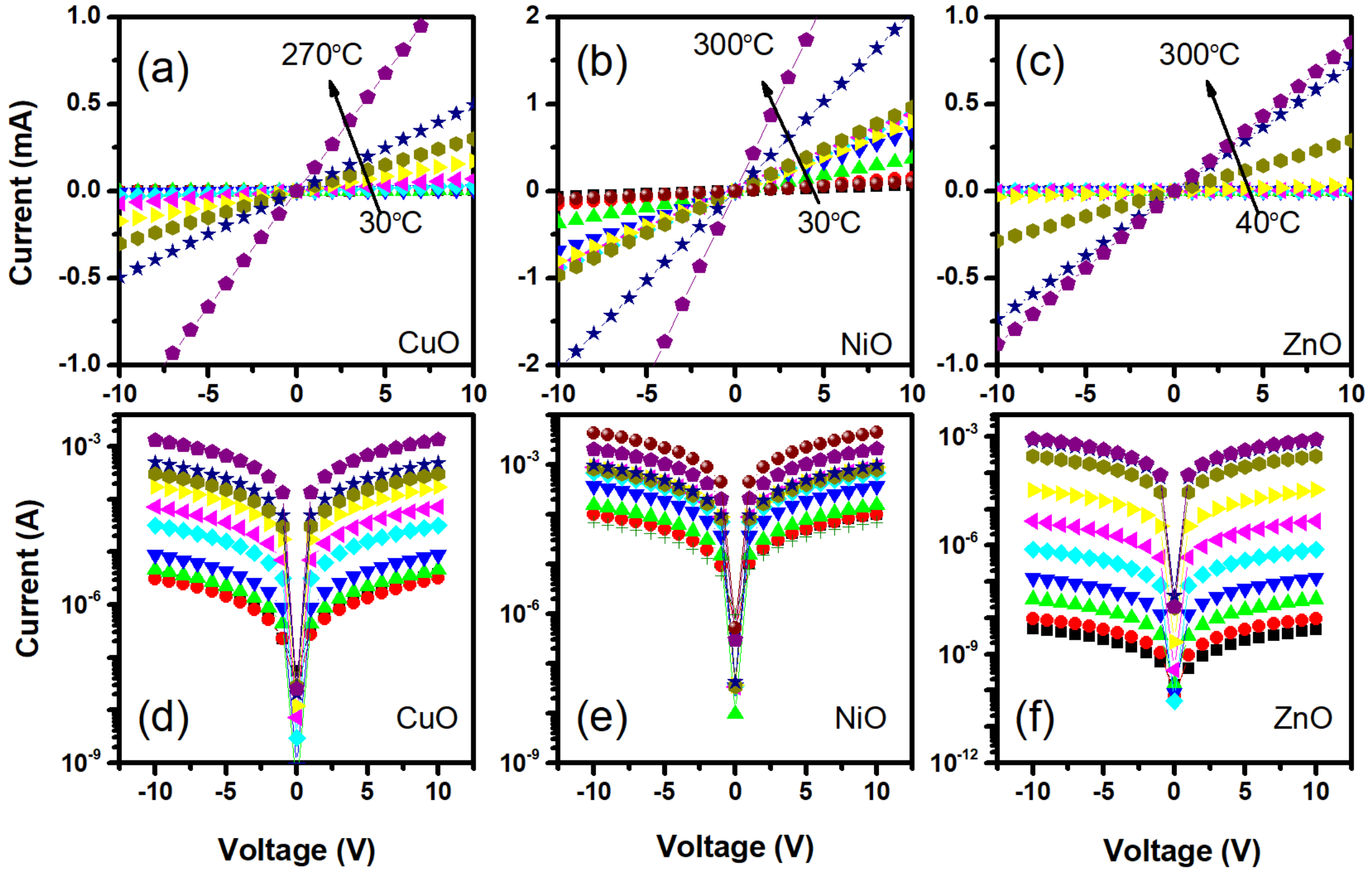

ESM 4: Current vs voltage plot for (a, d) CuO, (b,e) NiO and (c,f) ZnO at different temperatures in linear (top row) and log scale (bottom row).

The nature of the contact (i.e. ohmic or schottky) were tested before performing the sensing studies by measuring I-V characteristics. The same are shown in ESM 4. The IVs have been found to be linear implying an ohmic contact. Besides, their temperature dependence is studied by taking IV at each temperature interval of 25 degrees from room temperature to 300 degrees.

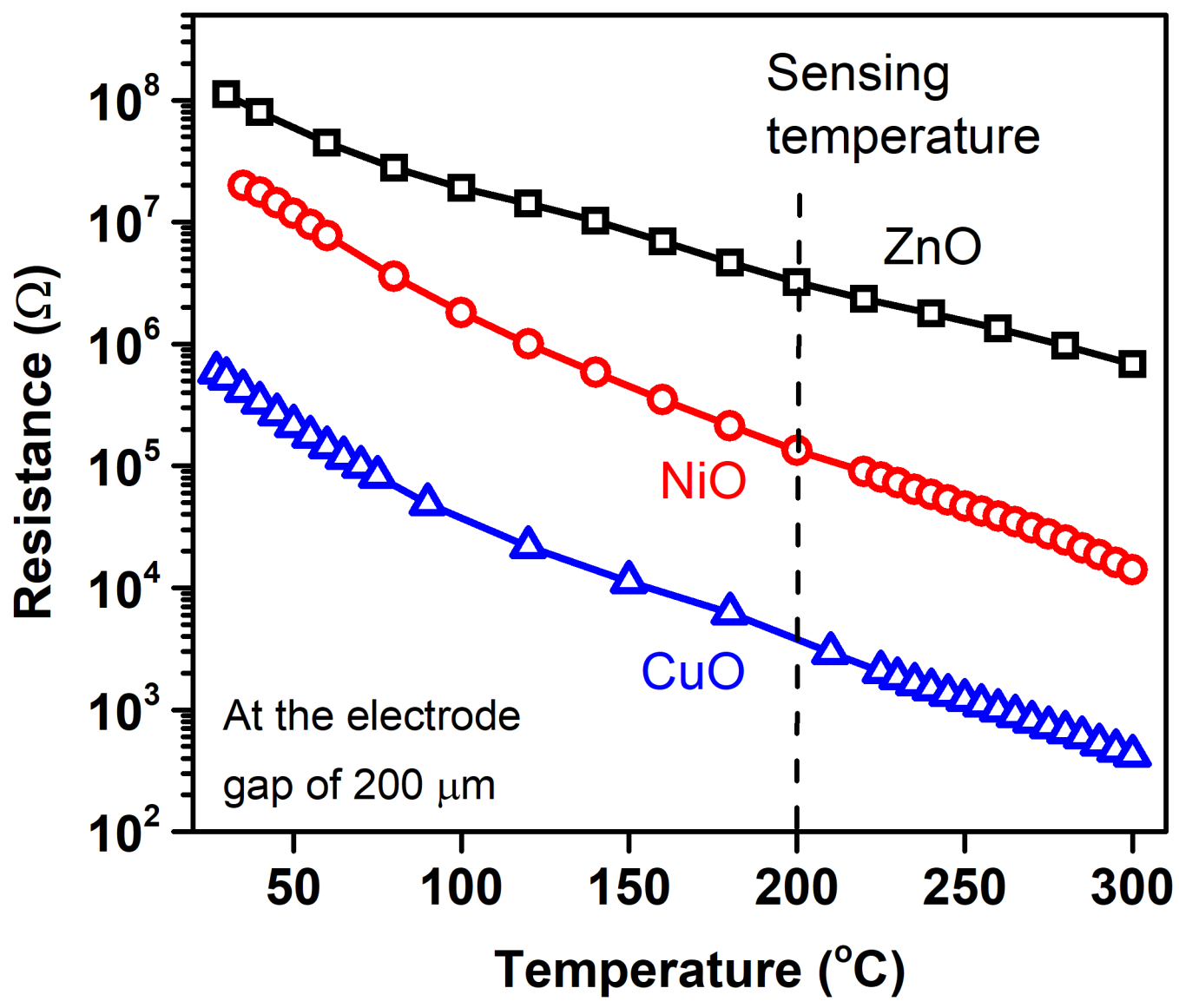

ESM 5: The Resistance vs. Temperature plot for all three metal oxides.

ESM 5: show the resistance vs temperature data of all the three sensor materials studied. They show a semiconducting behavior as expected. Among all the three sensors ZnO is the highest resistance sample. Nevertheless, at operating conditions of 200 °C all the three resistance are below 10 MΩ value which is suitable for the electronics.

## Gas sensing studies.

The gas response for each of the studies undertaken to study the response of single, double, and 3-gas simultaneously have been shown below. The data shown here has been used for the analysis mentioned in the main text.

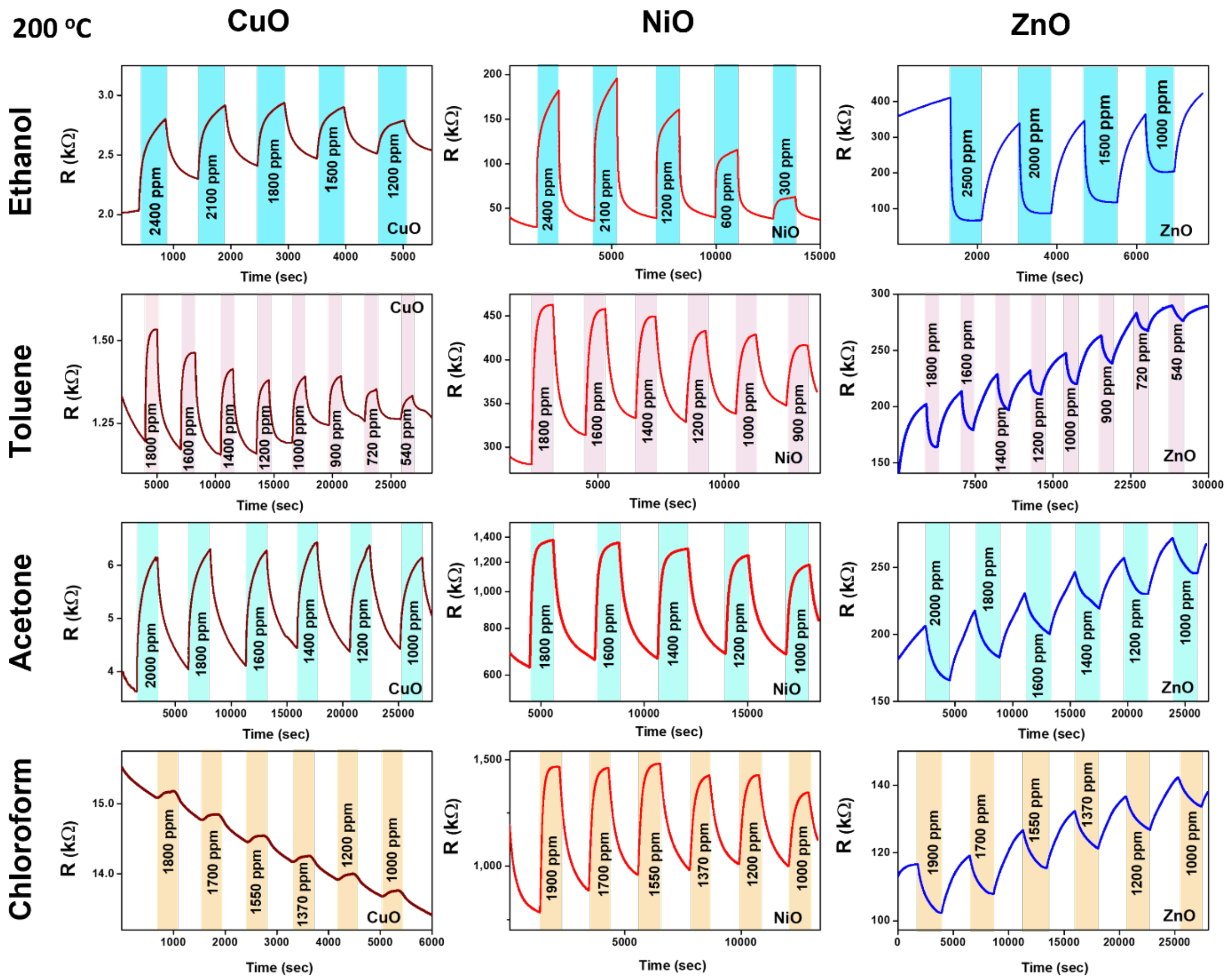

ESM 6: The Resistance vs time plots of CuO, NiO and ZnO for Ethanol, Toluene, Acetone, and Chloroform respectively. (single gas)

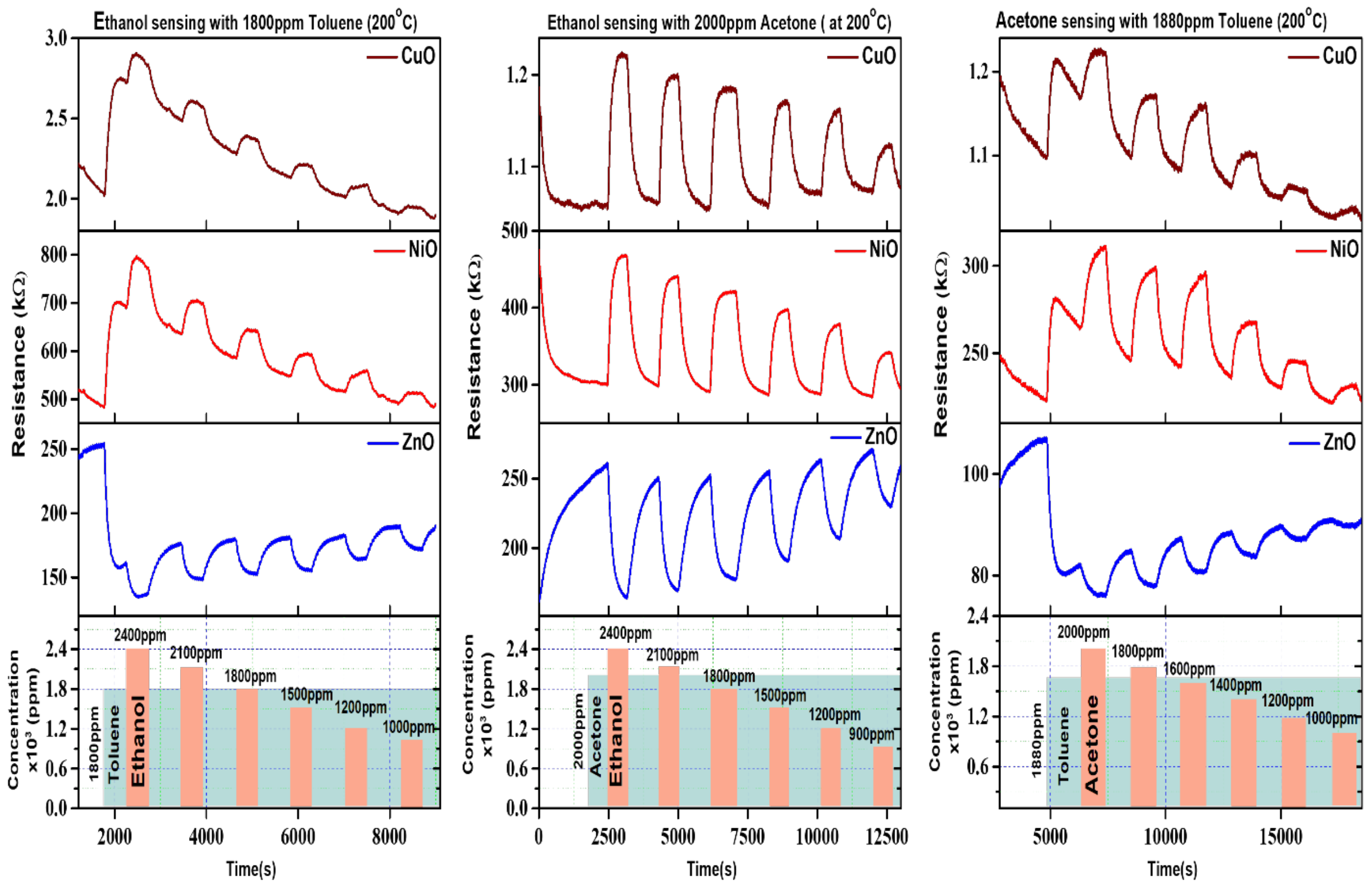

ESM 7: The double gas response was studied with one stationary and one variable gas. (two gas)

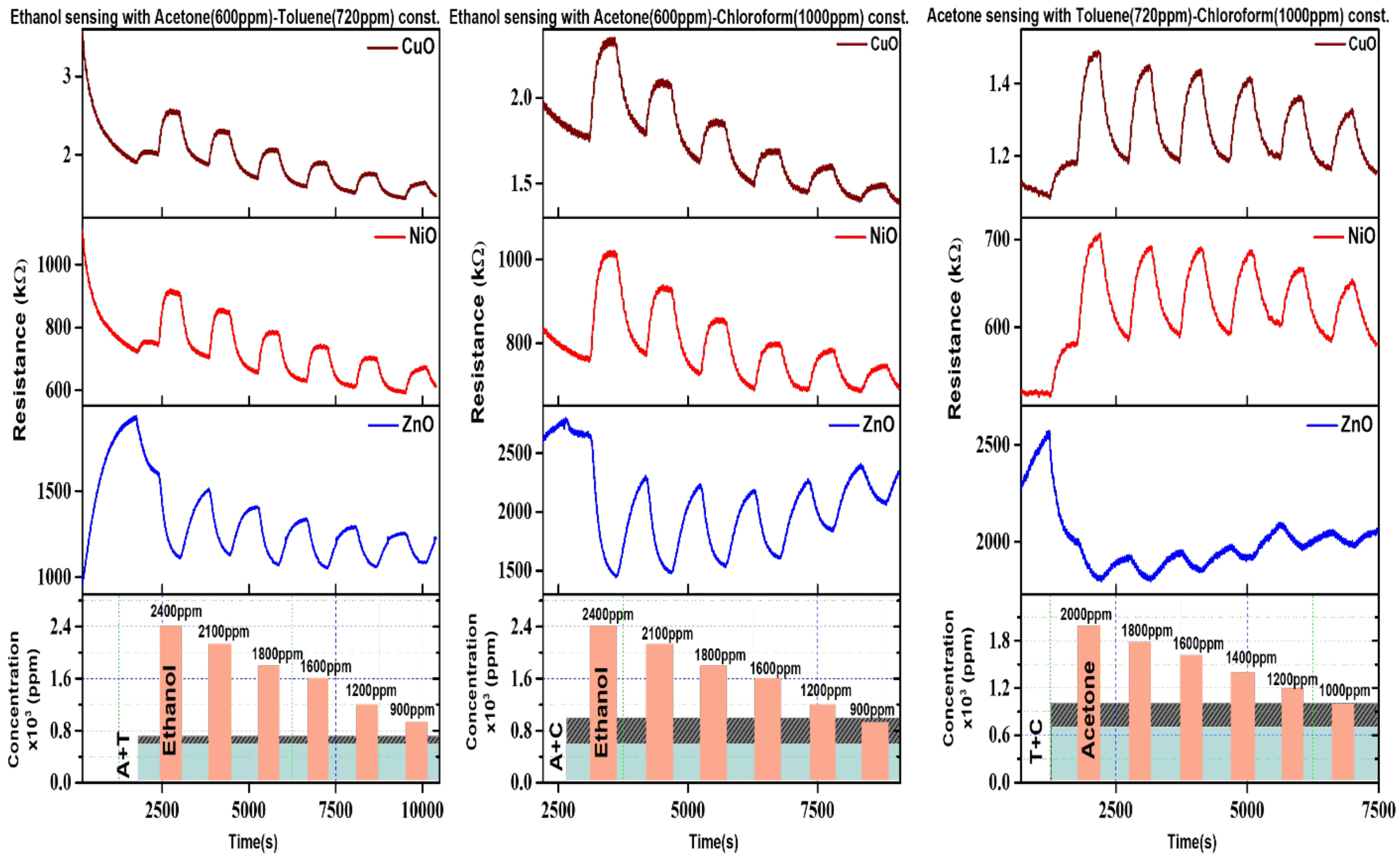

ESM 8. The response to ethanol vapors in presence of mixtures of toluene, acetone and chloroforms. (three gas)

The gas sensor response of CuO and ZnO for all the four gases individually as well as with the interfering species has been studied and the same is shown in ESM 8. It may be seen that the response is significantly different when measured individually vis-à-vis in presence of other gases.

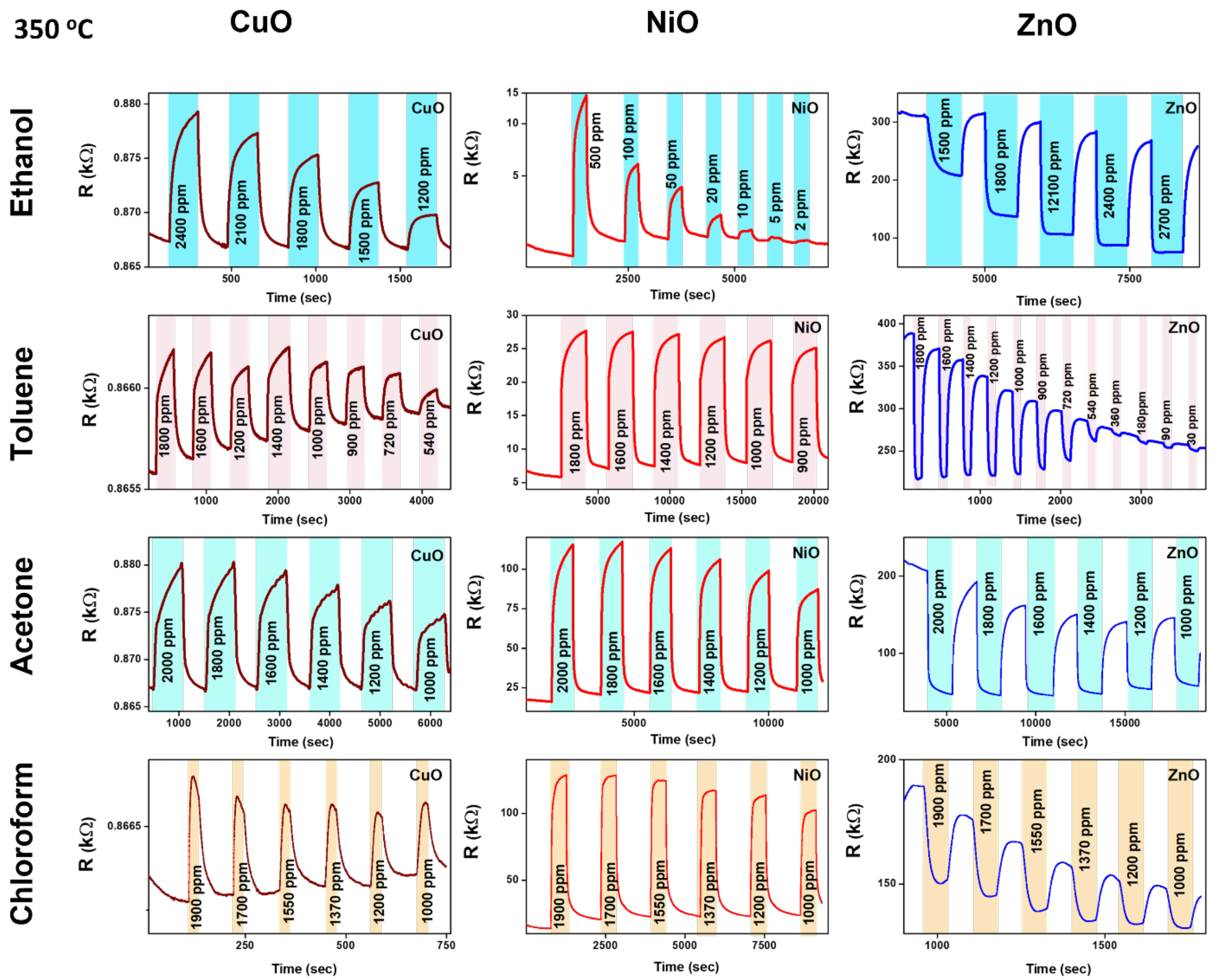

ESM 9: The Resistance vs. time plots of CuO, NiO, and ZnO for Ethanol, Toluene, Acetone, and Chloroform, respectively, at 350°C. (single gas)

Table S3: Variances of first 5 principal components (PC) over various gas mixture datasets

| | PC / Dataset | PC1 | PC2 | PC3 | PC4 | PC5 |
|---|---|---|---|---|---|---|
| **Variability (%)** | **1-gas** | 38.81186 | 28.86281 | 20.33076 | 8.01912 | 3.97542 |
| | **2-gases** | 41.81045 | 23.51523 | 15.22549 | 8.58879 | 7.50794 |
| | **3-gases** | 34.83254 | 23.81192 | 17.54433 | 10.54873 | 5.85302 |

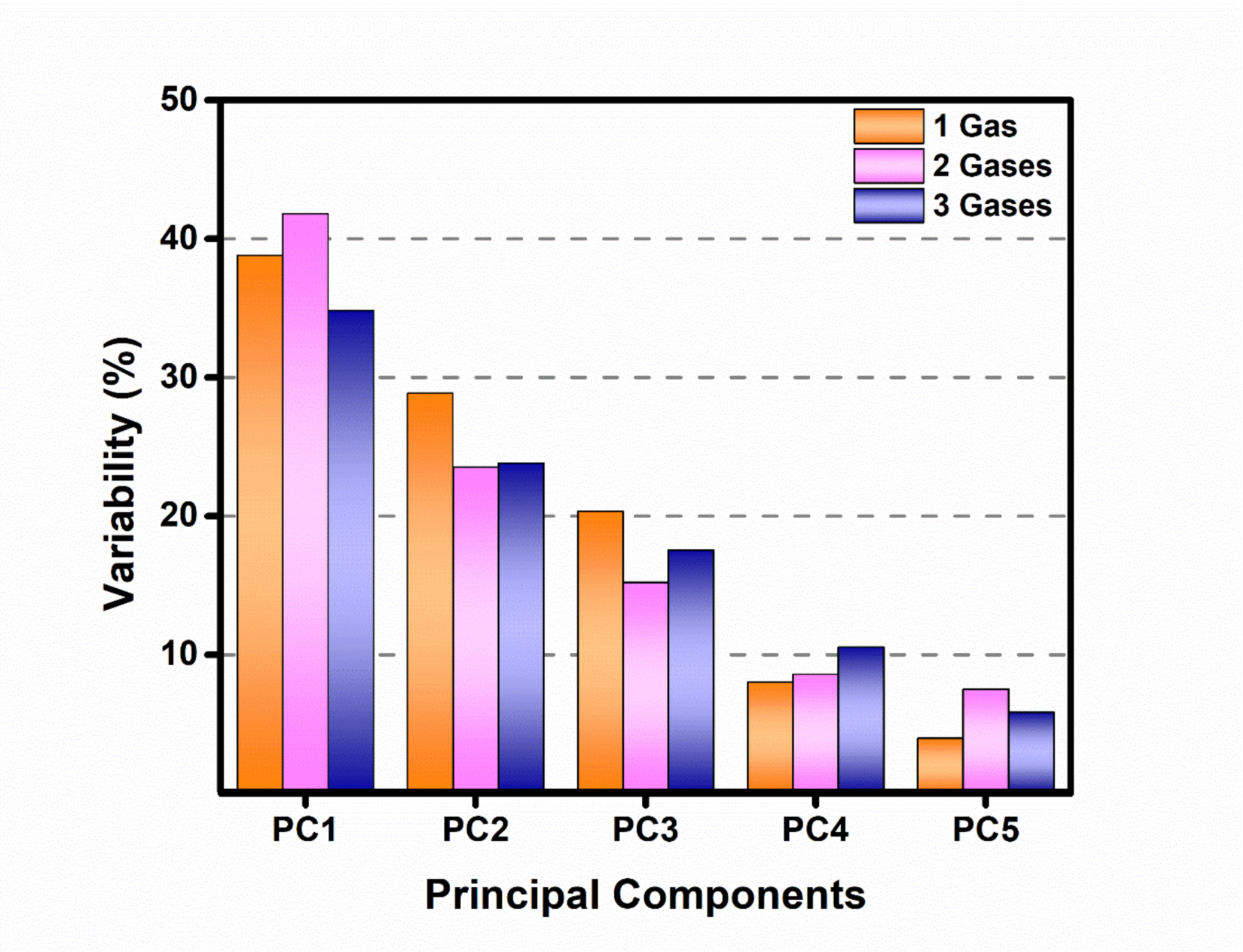

ESM 10: Pictorial representation of variability of first 5 principal components (PC) over various gas mixture datasets

To reduce the complexity of the data while preserving trends and patterns, we used Principal Component Analysis (PCA) on the sensor signal response. The variances of first 5 principal components (PC1, PC2, PC3, PC4, and PC5) are shown in Table S3 for 1-gas, 2-gases, and 3-gases datasets, and also represented pictorially in Fig. S11.

Considering PC1 and PC2, we obtained the 2D plots of ESM 11 and ESM 12 over 1-gas and 2-gases datasets, respectively.

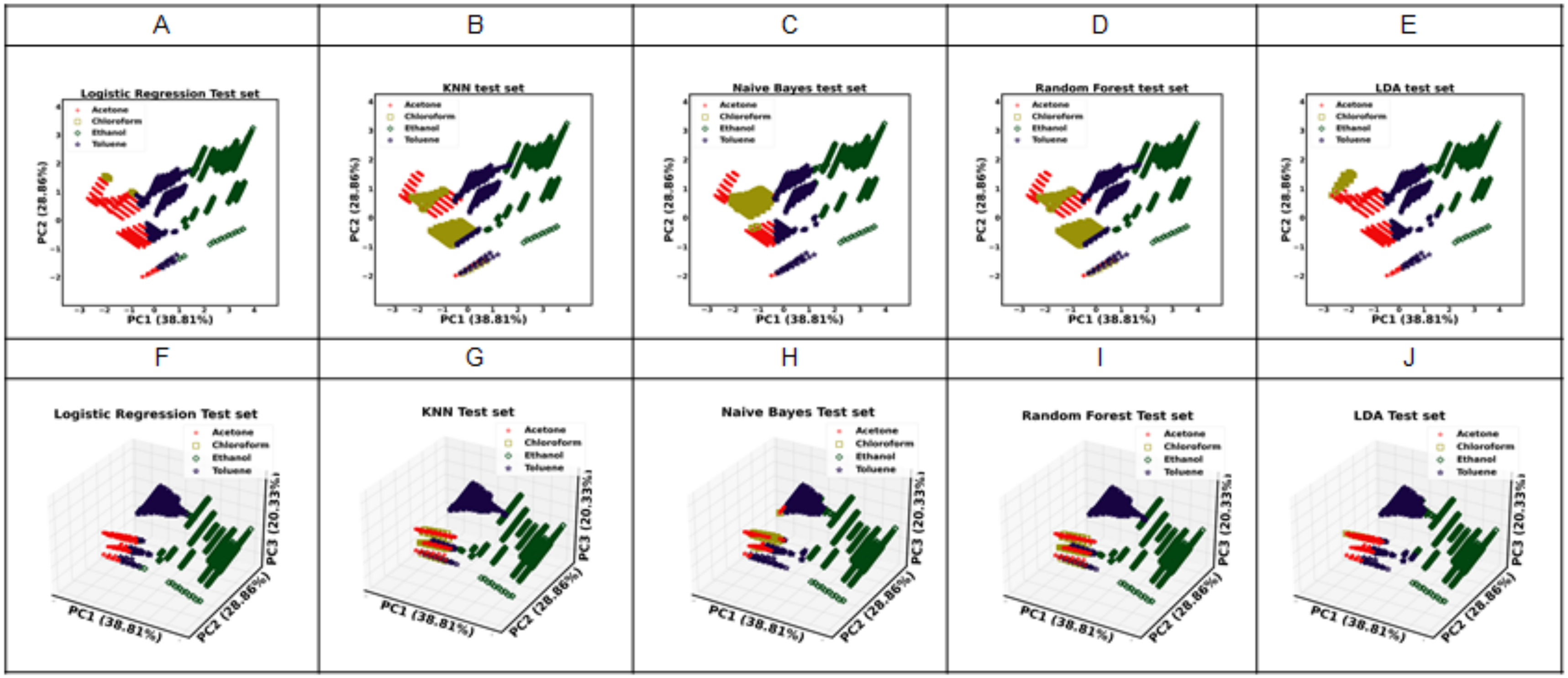

ESM 11: 1-gas dataset: 2D and 3D classification plots in 1st and 2nd rows, respectively.

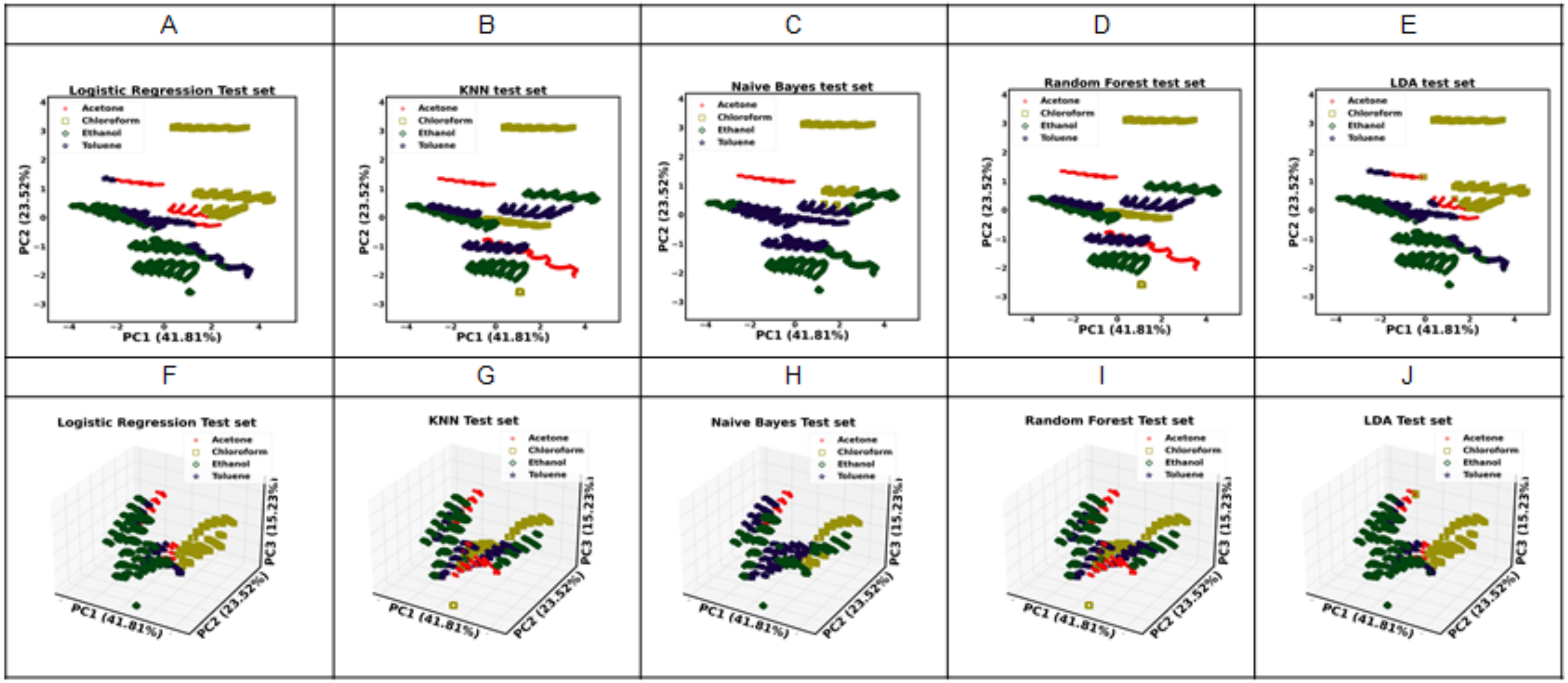

ESM 12: 2-gases dataset: 2D and 3D classification plots in 1st and 2nd rows, respectively.

Tables S4, S5, and S6 compare the experimental results obtained on 1-gas, 2-gases, and 3-gases datasets using KNN regression, ANN, random forest, decision tree, and linear regression models.

Table S4: Applying various machine learning model 1-gas dataset

| Method | Gas Name | RMSE | MSE | MAE | NRMSE | $R^2$ | LoD | LoQ |
|---|---|---|---|---|---|---|---|---|
| **KNN Regression** | Acetone | 0.00086 | $7.43\times10^{-7}$ | 0.00001 | 0.00114 | 0.99997 | 0.00344 | 0.01146 |
| | Toluene | 0.00082 | $6.77\times10^{-7}$ | 0.00001 | 0.00109 | 0.99997 | 0.00328 | 0.01095 |
| | Ethanol | 0.00076 | $5.82\times10^{-7}$ | 0.00001 | 0.00101 | 0.99997 | 0.00304 | 0.01015 |
| | Chloroform | 0.00153 | $2.35\times10^{-6}$ | 0.00004 | 0.00203 | 0.99990 | 0.00611 | 0.02039 |
| **ANN** | Acetone | 0.23798 | 0.05663 | 0.17747 | 0.58080 | 0.43819 | 1.83359 | 6.11199 |
| | Toluene | 0.37517 | 0.14075 | 0.30339 | 0.67179 | 0.00000 | 6.86907 | 22.89691 |
| | Ethanol | 0.46381 | 0.21512 | 0.37303 | 0.82863 | 0.00000 | 7.55353 | 25.17844 |
| | Chloroform | 0.02368 | 0.00056 | 0.00787 | 0.04346 | 0.99435 | 0.13167 | 0.43890 |
| **Random Forest** | Acetone | 0.01235 | 0.00015 | 0.00248 | 0.01645 | 0.99412 | 0.05032 | 0.16774 |
| | Toluene | 0.01300 | 0.00016 | 0.00265 | 0.01730 | 0.99348 | 0.05299 | 0.17663 |
| | Ethanol | 0.01218 | 0.00014 | 0.00233 | 0.01621 | 0.99426 | 0.04955 | 0.16519 |
| | Chloroform | 0.01215 | 0.00014 | 0.00233 | 0.01618 | 0.99428 | 0.04945 | 0.16484 |
| **Decision Tree** | Acetone | 0.08577 | 0.00735 | 0.06750 | 0.11416 | 0.71512 | 0.47851 | 1.59505 |
| | Toluene | 0.08602 | 0.00739 | 0.06766 | 0.11452 | 0.71391 | 0.48104 | 1.60349 |
| | Ethanol | 0.08589 | 0.00737 | 0.06761 | 0.11437 | 0.71547 | 0.47996 | 1.59987 |
| | Chloroform | 0.08596 | 0.00738 | 0.06762 | 0.11444 | 0.71526 | 0.48053 | 1.60176 |
| **Linear Regression** | Acetone | 0.06823 | 0.00465 | 0.05018 | 0.09082 | 0.82034 | 0.33194 | 1.10646 |
| | Toluene | 0.06830 | 0.00466 | 0.05022 | 0.09095 | 0.81937 | 0.33265 | 1.10883 |
| | Ethanol | 0.06839 | 0.00467 | 0.05033 | 0.09102 | 0.81945 | 0.33327 | 1.11091 |
| | Chloroform | 0.06827 | 0.00466 | 0.05025 | 0.09087 | 0.82019 | 0.33250 | 1.10833 |

Table S5: Applying various machine learning model 2-gases dataset

| Method | Gas Name | RMSE | MSE | MAE | NRMSE | $R^2$ | LoD | LoQ |
|---|---|---|---|---|---|---|---|---|
| **KNN** | Acetone | 0.00131 | $1.72\times10^{-6}$ | 0.00002 | 0.00319 | 0.99996 | 0.00957 | 0.03190 |
| | Toluene | 0.00094 | $8.98\times10^{-7}$ | 0.00001 | 0.00226 | 0.99998 | 0.00678 | 0.02260 |
| | Ethanol | 0.00095 | $9.21\times10^{-7}$ | 0.00001 | 0.00230 | 0.99998 | 0.00692 | 0.02309 |
| | Chloroform | 0.00194 | $3.79\times10^{-6}$ | 0.00006 | 0.00466 | 0.99992 | 0.01400 | 0.04669 |
| **ANN** | Acetone | 0.29005 | 0.08413 | 0.18941 | 0.74214 | 0.24191 | 4.13273 | 13.77579 |
| | Toluene | 0.26750 | 0.07155 | 0.22071 | 0.70174 | 0.39794 | 3.74026 | 12.46753 |
| | Ethanol | 0.29169 | 0.08508 | 0.23761 | 0.61701 | 0.21728 | 4.07928 | 13.59760 |
| | Chloroform | 0.01832 | 0.00033 | 0.00732 | 0.03550 | 0.99656 | 0.10766 | 0.35887 |
| **Random Forest** | Acetone | 0.00197 | $3.89\times10^{-6}$ | 0.00018 | 0.00476 | 0.99992 | 0.01429 | 0.04766 |
| | Toluene | 0.00182 | $3.33\times10^{-6}$ | 0.00017 | 0.00437 | 0.99993 | 0.01313 | 0.04378 |
| | Ethanol | 0.00108 | $1.18\times10^{-6}$ | 0.00009 | 0.00260 | 0.99997 | 0.00782 | 0.02608 |
| | Chloroform | 0.00182 | $3.35\times10^{-6}$ | 0.00018 | 0.00441 | 0.99993 | 0.0132 | 0.04414 |
| **Decision Tree** | Acetone | 0.07637 | 0.00583 | 0.06633 | 0.18478 | 0.88192 | 0.62754 | 2.09183 |
| | Toluene | 0.07653 | 0.00585 | 0.06650 | 0.18394 | 0.88135 | 0.62767 | 2.09226 |
| | Ethanol | 0.07667 | 0.00587 | 0.06653 | 0.18448 | 0.87986 | 0.62808 | 2.09362 |
| | Chloroform | 0.07677 | 0.00589 | 0.06683 | 0.18403 | 0.88072 | 0.62697 | 2.08992 |
| **Linear Regression** | Acetone | 0.03360 | 0.00112 | 0.02753 | 0.08087 | 0.97698 | 0.24889 | 0.82965 |
| | Toluene | 0.03379 | 0.00114 | 0.02768 | 0.08132 | 0.97672 | 0.24978 | 0.83262 |
| | Ethanol | 0.03351 | 0.00112 | 0.02747 | 0.08055 | 0.97710 | 0.24709 | 0.82363 |
| | Chloroform | 0.03366 | 0.00113 | 0.02756 | 0.08109 | 0.97695 | 0.24957 | 0.83193 |

Table S6: Applying various machine learning model 3-gases dataset

| Method | Gas Name | RMSE | MSE | MAE | NRMSE | $R^2$ | LoD | LoQ |
|---|---|---|---|---|---|---|---|---|
| **KNN** | Acetone | 0.00163 | $2.67\times10^{-6}$ | 0.00005 | 0.00393 | 0.99994 | 0.01179 | 0.03932 |
| | Toluene | 0.00204 | $4.19\times10^{-6}$ | 0.00006 | 0.00496 | 0.99991 | 0.01488 | 0.04961 |
| | Ethanol | 0.00196 | $3.87\times10^{-6}$ | 0.00005 | 0.00474 | 0.99992 | 0.01422 | 0.04742 |
| | Chloroform | 0.00342 | $1.17\times10^{-5}$ | 0.00020 | 0.00825 | 0.99976 | 0.02478 | 0.08260 |
| **ANN** | Acetone | 0.12812 | 0.01641 | 0.10564 | 0.23949 | 0.85374 | 0.82988 | 2.76628 |
| | Toluene | 0.25066 | 0.06283 | 0.19068 | 0.64111 | 0.48192 | 2.54640 | 8.48802 |
| | Ethanol | 0.18948 | 0.03590 | 0.15538 | 0.32139 | 0.66963 | 1.31273 | 4.37577 |
| | Chloroform | 0.03286 | 0.00108 | 0.01765 | 0.05478 | 0.98816 | 0.16329 | 0.54430 |
| **Random Forest** | Acetone | 0.00365 | $1.34\times10^{-5}$ | 0.00016 | 0.00883 | 0.99972 | 0.02650 | 0.08836 |
| | Toluene | 0.00273 | $7.46\times10^{-6}$ | 0.00010 | 0.00661 | 0.99984 | 0.01985 | 0.06616 |
| | Ethanol | 0.00227 | $5.19\times10^{-6}$ | 0.00007 | 0.00548 | 0.99989 | 0.01646 | 0.05488 |
| | Chloroform | 0.00369 | $1.36\times10^{-5}$ | 0.00016 | 0.00882 | 0.99972 | 0.02648 | 0.08828 |
| **Decision Tree** | Acetone | 0.05921 | 0.00350 | 0.04852 | 0.14292 | 0.92822 | 0.46004 | 1.53348 |
| | Toluene | 0.05986 | 0.00358 | 0.04902 | 0.14489 | 0.92787 | 0.46872 | 1.56241 |
| | Ethanol | 0.05983 | 0.00358 | 0.04903 | 0.14388 | 0.92846 | 0.46418 | 1.54728 |
| | Chloroform | 0.06019 | 0.00362 | 0.04923 | 0.14590 | 0.92783 | 0.47167 | 1.57224 |
| **Linear Regression** | Acetone | 0.03552 | 0.00126 | 0.02762 | 0.08468 | 0.97455 | 0.26122 | 0.87076 |
| | Toluene | 0.03508 | 0.00123 | 0.02728 | 0.08471 | 0.97506 | 0.25984 | 0.86614 |
| | Ethanol | 0.03550 | 0.00126 | 0.02764 | 0.08538 | 0.97480 | 0.26317 | 0.87725 |
| | Chloroform | 0.03531 | 0.00124 | 0.02743 | 0.08538 | 0.97501 | 0.26287 | 0.87623 |

Table S7: Comparison of $R^2$ obtained by employed ML-based regression architectures

| Dataset | Gas Name | KNN Regression | ANN | Random Forest | Decision Tree | Linear Regression |
|---|---|---|---|---|---|---|
| **1-gas** | Acetone | **0.99997** | 0.43819 | 0.99412 | 0.71512 | 0.82034 |
| | Toluene | **0.99997** | 0.00000 | 0.99348 | 0.71391 | 0.81937 |
| | Ethanol | **0.99997** | 0.00000 | 0.99426 | 0.71547 | 0.81945 |
| | Chloroform | **0.99990** | 0.99435 | 0.99428 | 0.71526 | 0.82019 |
| **2-gases** | Acetone | **0.99996** | 0.24191 | 0.99992 | 0.88192 | 0.97698 |
| | Toluene | **0.99998** | 0.39794 | 0.99993 | 0.88135 | 0.97672 |
| | Ethanol | **0.99998** | 0.21728 | 0.99997 | 0.87986 | 0.97710 |
| | Chloroform | 0.99992 | 0.99656 | **0.99993** | 0.88072 | 0.97695 |
| **3-gases** | Acetone | **0.99994** | 0.85374 | 0.99972 | 0.92822 | 0.97455 |
| | Toluene | **0.99991** | 0.48192 | 0.99984 | 0.92787 | 0.97506 |
| | Ethanol | **0.99992** | 0.66963 | 0.99989 | 0.92846 | 0.97480 |
| | Chloroform | **0.99976** | 0.98816 | 0.99972 | 0.92783 | 0.97501 |

*email: chandranath@iitp.ac.in ; r.p.shukla@utwente.nl ; kbvinayak@iisertvm.ac.in